\documentclass{article}

\usepackage{algorithm,algpseudocode}
\usepackage{algorithm}
\usepackage[noEnd=true]{algpseudocodex}

\usepackage{hyperref}

\usepackage{amssymb}
\usepackage{amsmath}
\usepackage{amsthm}

\usepackage{hyperref} 
\usepackage{url}

\usepackage{boxedminipage}
\usepackage{graphicx}
\usepackage{float}
\usepackage{rotating}

\usepackage{xspace}
\usepackage{nicefrac}

\usepackage{comment}
\usepackage{soul}
\usepackage{verbatim}

\usepackage{csquotes} 

\usepackage[T1]{fontenc}    
\usepackage[utf8]{inputenc} 
\usepackage{xparse}   
\usepackage{xifthen} 
\usepackage{pifont}
\usepackage{marvosym}

\usepackage{fullpage}
\usepackage{placeins}
\usepackage[most]{tcolorbox}

\newif\ifSLIDES
\SLIDESfalse

\newif\ifLONG
\LONGfalse

\newif\ifTWOCOLUMNS
\TWOCOLUMNSfalse

\newif\ifVERSIONCONF
\VERSIONCONFfalse

\input{macros-abs.sty}
\input{macros-pap-sesc.sty}

\newcommand{\SCgenericity}{
\subsection{Genericity assumptions}
\label{sec:genericity}

The genericity condition for our algorithm to work at a point $x$ is
that the intersection of spheres containing $x$ is transverse. For $k$
spheres, this existence of a non transverse intersection is checked as
follows.  The non transversality at point $x$ reads as $\sum_i
\lambda_i (x-c_i)=0$ or equivalently $(\sum_i \lambda_i)x = \sum_i
\lambda_i c_i$, which requires discussing two cases: (Case 1) $\sum_i
\lambda_i \neq 0$, and (Case 2) $\sum_i \lambda_i = 0$. In each case,
we need to check the points $x$ satisfying these conditions satisfy
the sphere equations.

To do so, Case 1 requires solving a QP problem. Case 2 requires
computing the null space of the matrix 
$A=( \{[c_i\ 1]^T\}_{i=1,\dots,k} )$, and if $Ker(A) \neq {0}$, one needs
to further check the sphere feasibility conditions, which require
solving another linear system.

Also note that at a given point $x$, the test simply boils down to
checking that the vectors $x-c_i$ are linearly independent.
\medskip

The exact computation of trajectories by our algorithm is more
involved. It requires cascaded degree two and degree four algebraic
numbers (degree two when intersecting a segment with a sphere; degree
four when intersecting a geodesic along a sphere with another sphere).

A robust numerical solution could be obtained using say an interval
number type with bounds of arbitrary precision, e.g. the iRRAM
library~\cite{muller2001irram}.

In practice though:
\begin{itemize}
\item We do not check the transversality condition, as even in medium
  dimensional spaces, the points where the intersections are not
  transverse are scarce, and our trajectories do not cross them.
\item We do not use elaborate number types, since the observed
  robustness of our floating point implementation did not require
  using them.
\end{itemize}
}

\title{Modeling high dimensional point clouds with the spherical cluster model}

\author{
Fr\'ed\'eric Cazals 
\thanks{\ucainria, France \texttt{frederic.cazals@inria.fr}},
Antoine Commaret \thanks{\ucainria, France \texttt{antoine.commaret@inria.fr}},
Louis Goldenberg
\thanks{Ecole Polytechnique, Department of Computer Science Palaiseau, France}
}

\begin{document}
\maketitle


\begin{abstract}
A parametric cluster model is a statistical model providing
geometric insights onto the points defining a cluster.
The {\em spherical cluster model} (SC) approximates a
finite point set $P\subset \mathbb{R}^d$ by a sphere $S(c,r)$ as follows.  Taking $r$
as a fraction $\eta\in(0,1)$ (hyper-parameter) of the std deviation of
distances between the center $c$ and the data points, the cost of the
SC model is the sum over all data points lying outside the sphere $S$
of their power distance with respect to $S$.
The center $c$ of the SC model is the point minimizing this cost.
Note that $\eta=0$ yields the celebrated center of mass used in
KMeans clustering. We make three contributions.

First, we show fitting a spherical cluster yields a strictly convex
but not smooth combinatorial optimization problem. Second, we present
an exact solver using the Clarke gradient on a suitable stratified
cell complex defined from an arrangement of hyper-spheres. Finally,
we present experiments on a variety of datasets
ranging in dimension from $d=9$ to $d=10,000$, with two main observations.
First, the exact algorithm is orders of magnitude faster than
BFGS based heuristics for datasets of small/intermediate dimension and
small values of $\eta$, and for high dimensional datasets (say
$d>100$) whatever the value of $\eta$. Second, the center of the SC
model behave as a parameterized high-dimensional median.

The SC model is of direct interest for high dimensional multivariate data analysis,
and the application to the design of mixtures of SC will be reported
in  a companion paper.
\end{abstract}


\section{Introduction}

\ifLONG
\else
\paramini{Clustering methods.}  Clustering, namely the task
which consists in grouping data items into dissimilar groups of
similar elements, is a fundamental problem in data analysis at
large~\cite{xu2005survey}.  Existing clustering methods may be
ascribed to four main tiers, including hierarchical
clustering~\cite{rod-peh-pcsa-73}, density based
clustering \cite{cheng1995mean,chazal2013persistence}, spectral
clustering~\cite{von2007tutorial}, as well as k-means
~\cite{arthur2007k} and variants related to
Expectation-Maximization \cite{dempster1977maximum,kasarapu2015minimum}.
The goal of clustering being to aggregate points into coherent groups,
equally important are {\em cluster models} aiming at providing
insights into the geometry of the data.  This goal prompted the
development of \ksubspace clustering techniques, which belong to two
tiers.
Sparse subspace clustering
(SSC/ASSC)~\cite{elhamifar2013sparse,soltanolkotabi2012geometric,li2018geometric}
reconstruct points as (sparse) linear combinations of neighbors, while
parametric cluster model use explicit analytical form for
clusters~\cite{parsons2004subspace,wang2009ksubspaces}.
\fi


\paramini{Geometric medians.}
Cluster models providing insights on the geometry of a point
set also call for a discussion of high dimensional medians.
The Fermat-Weber point is the point from $\Rd$ minimizing the sum of
Euclidean distances to all data points.  Unfortunately, this point is
hard to compute and unstable
\cite{kupitz1997geometric,weiszfeld1937point,bajaj1988algebraic,cohen2016geometric}.
Building on Helly's theorem, a median can be defined as any point
whose Tukey depth is at least $\geq n/(d+1)$~\cite{tukey1975mathematics}.
\ifLONG
\else
Such a point is, however, hard to compute 
\cite{clarkson1993approximating,miller2009approximate}.
\fi
The projection median is defined by projecting the dataset onto random
lines, computing the univariate median for each projection, and
computing a weighted average of the data points responsible for these
univariate
medians~\cite{durocher2009projection,basu2012projection,durocher2017projection}.
It is an elegant, stable and remarkably effective generalization of
the univariate median.

\paramini{Contributions.}  Two types of parametric cluster models have
been proposed recently~\cite{wang2009ksubspaces}: affine and spherical
clusters (SC). The former accommodates potentially unbounded (large)
clusters of arbitrary dimension.  The latter defines compact
(spherical) clusters based on the power distance of points with
respect to a sphere whose radius is (a fraction of) the variance of
distances to the cluster center--to be determined.
However, the uniqueness of the SC center is not established,
and no algorithm is presented to compute it. (The calculation
presented assumes the center is known, and it solely observes that the
result obtained is consistent with the usual center of mass when the
fraction of the variance tends to zero~\cite{wang2009ksubspaces}.)

We make two contributions. First, we establish the SC cluster model is well posed (the
solution is unique) and also well poised (numerically tractable).
Second, we present experiments showing that the center of the SC model
behave as a parameterized high-dimensional median.

The application of SC embedded into affine subspaces of positive
codimension will be presented in a companion paper.

All proofs and detailed algorithms are provided in the Supporting Information.

\section{Parametric cluster models and  Subspace Embedded Spherical Clusters} 
\label{sec:sesc}

\subsection{Notations}


Let $D$ be a set of $n$ points in $\Rd$.  We consider a partition of
$D$ into $k$ clusters $C_1, \dots, C_k$, with $D_\ell$ the set of
points associated to cluster $C_\ell$.
The unbiased variance estimate for distances within cluster $D_\ell$ of center $c_\ell$ satisfies
\begin{equation}
\label{eq:stdevdist}
\stdevdist* =  \frac{1}{n-1}\sum_{x_i\in D_\ell} \vvnorm{x_i - c_\ell}^2
\end{equation}

Let $A = c + V$ be an affine space, with $c$ a
point in $\Rd{d}$ (think cluster center), and $V$ a vector space.  For any
point $x \in \Rd$, we denote by $\comppara{x-c}$ the orthogonal
projection of the vector $(x-c)$ onto $V$, and by $\compperp{x-c}$ the
orthogonal projection on $V^\perp$.

When fitting a model, the sum of squared distances from samples to the
model is called the {\em residual sum of squares (RSS)}, or dispersion
for short.

\subsection{Parametric cluster models}


Take $C_\ell$ for $\ell \in \intrange{1}{k}$ and suppose $D_\ell$ is
known. Cluster $C_\ell$ is described by the parameter set $\theta_\ell =
(\theta_{\ell,1}, ..., \theta_{\ell,r})$ and a function $d_\ell:
(x,C_\ell(\theta_\ell)) \mapsto d_\ell(x,C_\ell(\theta_\ell))$, that
is some distance from a point to the cluster. We call the description
of $C_\ell$ by the function $d_\ell$ a {\em parametric cluster model}.
We decompose the clustering problem into two sub-problems concerned with the minimization of
a  dispersion term based on squared distances:
\begin{problem}[Cluster optimization] 
\label{pb:cluster-optim}
Let  $C_\ell$ be a parametric cluster. {\em Cluster optimization} is the optimization
problem seeking the cluster parameters minimizing the dispersion 
\begin{equation}
\label{eqn:generaloptimsinglecluster}
\min_{\theta_\ell} \kmfunc, \text{ with } \kmfunc = \sum_{x \in D_\ell} d_\ell(x,C_\ell(\theta_\ell))^2.
\end{equation}
\end{problem}

\paramini{Affine clusters.}
As a first generalization of \kmeans, one can consider the distance from a data point
to an affine subspace, yielding \ksubspace clustering~\cite{wang2009ksubspaces}:
\begin{definition}[Subspace cluster] \label{def:subspacecluster}
Let $A = c + V$ be some affine subspace of $\Rd$ where $c \in
\Rd$ is a point and $V$ is an $m$-dimensional linear subspace. The
    {\em subspace cluster} $C_\ell(A)$ is a cluster, where the
    distance from a point $x$ to the cluster is the distance to the
    subspace :
\begin{equation}
\label{eqn:subspacedistance}
d(x, C_\ell(A))^2 := d(x,A)^2 = \vvnorm{ \compperp{ x-c}}^2.
\end{equation}
\end{definition}

\paramini{Spherical clusters.}
As noticed in Introduction, affine clusters may be confounded by
noise, and suffer from their non compact nature.  This latter aspect
can be taken care of using {\em spherical clusters}. To see how,
recall that the power of a point $x$ with respect to a sphere $S(c,r)$
is defined by $\powerps{x}{S} = \vvnorm{x-c}^2 - r^2$.
Following \cite{wang2009ksubspaces}, we define:
\begin{definition}[Spherical cluster] 
\label{def:sphericalcluster}
Let $\eta \in ]0,1[$ be is a hyperparameter, and let $c_\ell$ be a
    point called the {\em cluster center}.  Given the set $D_\ell$,
    the distance function associated to the {\em spherical cluster}
    $C_\ell(c_\ell)$ reads as
\ifTWOCOLUMNS
\begin{align}
\label{eqn:sphericaldistance}
\begin{split}
d(x,C_\ell(c_\ell))^2 
&:=  \max \left(0, \vvnorm{x-c_\ell}^2 - \eta \stdevdist*  \right)  \\
&= \max \left(0, \powerps{x}{S(c_\ell, \sqrt{\eta} \stdevdist)}\right).
\end{split}
\end{align}
\end{definition}
\else
\begin{equation}
\label{eqn:sphericaldistance}
d(x,C_\ell(c_\ell))^2 
:=  \max \left(0, \vvnorm{x-c_\ell}^2 - \eta \stdevdist*  \right)  = \max \left(0, \powerps{x}{S(c_\ell, \sqrt{\eta} \stdevdist)}\right).
\end{equation}
\end{definition}
\fi
The rationale of this definition is that one wishes to find the center
minimizing  the cost of outliers--points outside the spherical cluster.

\subsection{Subspace-embedded spherical cluster}

We note that {\em subspace-embedded spherical clusters}
can be defined for affine spaces of arbitrary co-dimension:
\begin{definition}[Subspace-embedded spherical cluster] 
\label{def:cluster}
Given a set of points $D_\ell$, we define a {\em subspace-embedded
  spherical cluster} as a spherical cluster of center $c$ embedded into 
an affine subspace $V$ of $\Rd$.
The corresponding distance  reads as
\ifTWOCOLUMNS
\begin{align}
\label{eqn:distance-sesc}
\begin{split}
& \dSESC^2 = \vvnorm{\compperp{x-c}}^2 +\\
& \mu \max \left(0, \vvnorm{\comppara{x-c}}^2 - \eta \frac{1}{n-1}\sum_{x_i\in D_\ell} \vvnorm{ \comppara{x_i-c}}^2 \right)
\end{split}
\end{align}
\else
\begin{equation}
\label{eqn:distance-sesc}
\dSESC^2 = 
\vvnorm{\compperp{x-c}}^2 + \mu \max \left(0, \vvnorm{\comppara{x-c}}^2 - \eta \frac{1}{n-1}\sum_{x_i\in D_\ell} \vvnorm{ \comppara{x_i-c}}^2 \right)
\end{equation}
\fi
\end{definition}
For a fixed data set $D_\ell$, the sum of the previous quantities
yields the dispersion to be minimized.

\section{Spherical Cluster Optimization}
\label{sec:sesc-opt}

This section studies problem \ref{pb:cluster-optim} for the spherical
cluster model (Def.  \ref{def:sphericalcluster}).  Since $\eta$ is fixed,
we simply denote $\Feta$ as $F$.

\subsection{Functional decomposition and geometry of the sub-functions}

For a
fixed data set $D_\ell$, we aim at minimizing over $\Rd$ the map 
\begin{equation}
\label{eq:Feta}
\ifTWOCOLUMNS
\small
\else
\fi
\Feta{c} := \sum_{x_i \in D_\ell} \max \left(0, \| x_i-c \|^2 - \frac{\eta}{n-1}\sum_{x_j\in D_\ell} \|x_j - c\|^2  \right).
\end{equation}
To study the previous function, for each $x_i \in D_\ell$, let
\begin{equation}
\ftildeeta{c}  := \| x_i-c \|^2 - \eta \frac{1}{n-1}\sum_{x_j\in D_\ell} \|x_j - c\|^2.
\end{equation}
so that 
\begin{equation}
\label{eqn:fetasum}
\Feta{c} = \sum_{x_i \in D_\ell} \max(0, f_{\eta,x_i}(c)).
\end{equation}
We first analyze the sub-functions and $f_{\eta,x_i}$ in order to
analyze the main function $\Feta$.  In the sequel, we assume that (i)
the set $D_\ell$ is fixed, (ii) $x_i \in D_\ell$, (iii) $0 < \eta < 1
- 1/n$ (iv) $\eta$ is fixed, so that we drop $\eta$ from the notations
(e.g writing $F, f_{x_i}$ instead of $F_{\eta}, f_{\eta, x_i}$ to ease
notations).

Studying the function $f_{x_i}$ benefits from the geometry of the
following {\em sink region} yielding a null cost: 
\begin{definition}
\label{def:sink}
The {\em sink region} $B_{x_i}$ is the set over which $f_{x_i}$ does
not contribute to $F$, that is $B_{x_i} := f_{x_i}^{-1}\left(- \infty,
0 \right]$. We denote $S_{x_i}$ its topological boundary.
\end{definition}
Remark that since $ \eta < 1 - \frac{1}{n}$ the intersection $B$ of all $B_{x_i}$ is necessarily empty, as any $x$ belong to this set would verify $\vvnorm{x - x_i}^2$ strictly lower than the average $\frac{1}{n} \sum_{x_i \in D_{\ell}} \vvnorm{x - x_i}^2$. \label{rk:emptyB}

The following results from an elementary calculation.
\begin{lemma}[Geometry of $B_{x_i}$] 
\label{lem:sink-geom}
Let $\eta' = \frac{n-1}{n} \eta$. Each map $f_{x_i}$ is proportional to a spherical power, and takes the form
\begin{equation}
f_{x_i}(c) = (1 - \eta') \left (\vvnorm{c - c_i}^2 - R_i^2 \right)
\end{equation} 
Putting $\bar{x} := \frac{1}{n} \sum_{x_j \in D_\ell} x_j$, the center $c_{i}$ and the radius $R_{i}$ of said sphere satisfy the following.
\begin{equation}
\label{eq:sink-spec}
\begin{cases}
c_{i} &= \frac{x_i - \eta'\bar{x}}{1-\eta'},\\
R_{i}^2 &= \left\| \frac{x_i - \eta'\bar{x}}{1-\eta'} \right\|^2 - \frac{\|x_i\|^2 - \frac{\eta'}{n} \sum_{x_j \in D_\ell} \|x_j\|^2}{1-\eta'} \bigr.  
\end{cases}
\end{equation}

As a consequence the sink region $B_{x_i}$ is a non-empty closed ball of $\Rd$, and $S_{x_i}$ is its associated sphere. 
\end{lemma}
As an immediate corollary, $\max(0, f_{x_i})$ is a convex map. In $\Rd \setminus B_{x_i}$, it is quadratic with gradient $\nabla f_{x_i}(c) = 2(1- \eta')(c - c_i)$, while being identically zero inside $B_{x_i}$.

\subsection{Arrangement of hyper-spheres underlying the objective function}


Our analysis of the sub-functions $f_{x_i}$ shows that they 
are continuous and {\em piecewise quadratic}: zero on $B_{x_i}$, and a quadratic form on $\Rd
\setminus B_{x_i}$. Thus, $F$ is also continuous, convex and
piecewise quadratic.
Finding the optimal cluster center requires
understanding the relationship between all sink regions.

\paramini{Arrangement.}
An {\em arrangement} of hyper-surfaces is a decomposition of 
$\Rd$ into equivalence classes of points using their position
with respect to these hyper-surfaces \cite{halperin2017arrangements}. 
We apply this concept to the spheres bounding the sink regions (Lemma
\ref{lem:sink-geom}).
For $x \in \Rd$ and $i \in \intrange{1}{n}$, consider the following
signature which states whether point $x$ lies outside/on/inside the
spheres $S_{x_i}$. It is a length $n$ vector with one entry in $\{-1, 0, 1 \}$ for each sink-defining ball:
\begin{equation}
    \sigma(x) := (sign(f_ {x_1}(x)), \dots,sign(f_ {x_n}(x))).
    \label{eqn:signature}
\end{equation}
The signature defines an equivalence relation, where two points are
equivalent if they have the same signature. We call {\em cells} the equivalence classes, and we use the notation $\mathcal{C}$ to denote them. By definition, cells are non-empty
and characterized by the three-set partition $I^+(\calC), I^0(\calC), I^-(\calC)$ of $\intrange{1}{n}$, where the sets are defined respectively as the sets of indices $i$ where $f_{x_i}$ are positive, zero, and negative. 
Note that generically, $\tau+1 \leq d$ spheres in dimension $d$
intersect along an $l=d-(\tau+1)$ sphere; thus we let the {\em
  dimension} of a cell $\calC$ be the number $d - \#I^0(\calC)$. Cells
of dimension $d$ are said to be {\em fully dimensional} and are open
subsets of $\Rd$, while others are said to be of \textit{positive
  codimension}.

\paramini{Combinatorial decomposition of $F$.}
On a cell $\calC$, $F$ is determined by the value of $f_{x_i}$ where $i$ ranges among $I^+(\calC)$. More precisely, we have
\begin{equation}
F_{|C}(c) = \sum_{i \in I^+(\calC)} f_{x_i}(c) = f_{I^+(\calC)}(c),
\end{equation}
where $f_{J} := \sum_{i \in J} f_{x_i}$ for any subset $J$ of $\intrange{1}{n}$. In the same vein, we put $S_J := \bigcap_{i \in J} S_{x_i}$ so that in a generic configuration of spheres any cell with non empty $I^0$ is a relatively open subset of $S_{I^0}$. We define $c_{J}$ to be the center of mass of all $c_i$ where $i$ ranges among $J$. 
\begin{equation}
\label{eq:Cj}
\optcell{J} :=  \frac{1}{\# J} \sum_{i \in J} c_i.
\end{equation}
Putting $R^2_J := \vvnorm{c_J}^2 + (\#J)^{-1} \left ( \sum_{i \in J} R^2_i - \vvnorm{c_i}^2 \right) $, straightforward computations yield
\begin{equation}
f_{J}(c) = (1 - \eta')\#J \left [ \vvnorm{c - c_J}^2 - R_J^2 \right].
\end{equation}
\toblack
Since any full-dimensional cell $\calC$ is open, $F$ is twice differentiable in $\calC$ with gradient and Hessian as follows:
\begin{equation}
\label{eqn:HessGradFullDim}
\begin{cases}
\nabla F_{|\calC}(c) = 2(1- \eta')\#I^+(\calC) \left( c - \optcell \right ) \\
H F_{|\calC}(c) = 2(1-\eta')\#I^+(\calC) \mathrm{Id}.
\end{cases}
\end{equation}


\subsection{Strict convexity and optimization}
\label{sec:strict-convex-optim}

The spherical cluster optimization is well-posed. Indeed we have seen that $B = \bigcap_{x_i \in D_{\ell}} B_{x_i}$ is empty, and from previous computations the Hessian of $F$ is almost-everywhere positive definite outside of $B$.

\begin{theorem}[Strict convexity of $F$] \label{thm:strcvxf}
Let $D_\ell$ be a set of $n$ points of $\Rd$, with at least two
distinct points, and assume that $0 < \eta < 1 - \frac{1}{n}$. 
Then the associated $F$ map is $2(1 - \eta')$-strongly convex on $\Rd$, and is a fortiori strictly convex.
Its minimization problem admits exactly one solution in $\Rd$.
\end{theorem}
\color{black}

\paramini{Minimum of $F$ on a full-dimensional cell.} From Eq. (\ref{eqn:HessGradFullDim}), it is clear that the minimum of $F$ is attained on a full-dimensional cell $\calC$ if and only if its gradient vanishes, translating to the following characterization.
\begin{equation}
\argmin[\Rd] F \in \calC \iff \optcell \in \calC.
\end{equation}

\begin{figure}[htb]
\begin{tabular}{cc}
\includegraphics[width=0.48\linewidth]{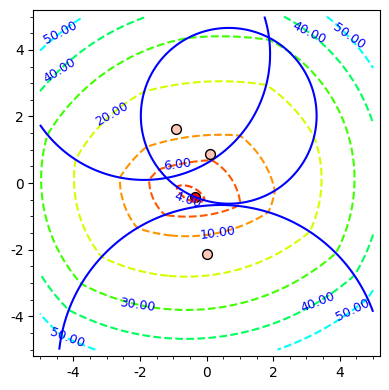}&
\includegraphics[width=0.48\linewidth]{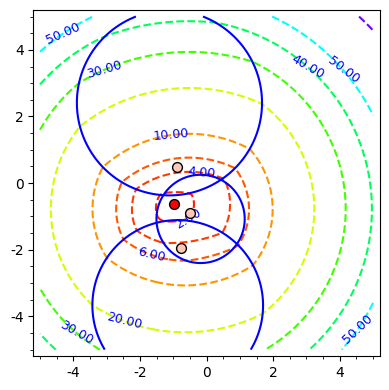}
\end{tabular}
\caption{{\bf Minima of $F$ on cells of various dimensions.}
Data point in orange, minima in red. Selected level sets (in dotted-lines) are also reported.}
\label{fig:opt-codim}
\end{figure}

To find the minimum of $F$, it is not enough to check all full-dimensional cells as it is absolutely possible that the minimum be attained on a cell with positive codimension  (see e.g.) Fig. \ref{fig:opt-codim}, left.), where no closed-form for a minimum can be obtained.

\section{Optimization: computing the unique minimizer of $\Feta$}
\label{sec:optim}

Having established the strict convexity of $\Feta$, 
we compute its unique minimizer. 
(As in the previous section, we simply denote 
$\Feta$ as $F$.)
Our algorithm actually minimizes  a function of the form $\sum_{i} \max(0, \vvnorm{x - c_i}^2 - R^2_i)$,
and constructs a finite sequence of points $(x_n)$ by induction. The last point is the optimum of $F$.
It assumes infinite precision -- the so called \textit{real RAM model},
and also assumes that all points $c_i$ are in a generic position and that when spheres $S_i$
intersect.
See Sections \ref{sec:numerics} and \ref{sec:genericity} for comments on these assumptions.


\subsection{Subdifferential and generalized gradient}

We face a non-smooth convex optimization problem without constraint. Guarantees on the speed of convergence of algorithms with such assumptions as a number of iterations are rather weak. While gradients exists almost-everywhere, the classical gradient descent method may get trapped to bottleneck situations \cite{Bubeck} leading to a precision rate of $O \left (\frac{1}{\sqrt{T}} \right)$, where $T$ is the number of steps. Further non-smooth investigation revolves around the use of the \textit{subdifferential} or equivalently the \textit{Clarke gradient} \cite{clarke1997nonsmooth} of the function. For a convex function, the subdifferential/Clarke gradient of $F$ at $x$, denoted by $\partial_* F(x)$ is a convex set defined below, while the generalized gradient $\nabla_* F(x)$ is its element of least norm.
\begin{equation}
\begin{cases}
\partial_* F(x) & :=  \{ s, f(y) - f(x) \geq s \cdot ( y - x), \forall y\} \\
\nabla_* F(x) & :=  \underset{u \in \partial_* F(x)} \argmin \vvnorm{u}. 
\end{cases}
\end{equation} 

Gradient samplings methods~\cite{gradient_sampling} avoid the earlier described bottleneck configurations with a good descent direction obtained by approximating the generalized gradient. Given an arrangement of the space, the recent so-called stratified gradient sampling \cite{stratified_gradient_sampling} proposes to use the arrangement to efficiently determine a good descent direction. To tailor an exact algorithm, we use the structure at our disposal. Indeed, for any fixed $x$ in $\Rd$, we let $I^+, I^0$ and $I^-$ be the three-set partition associated to the cell of $x$. 
Then the subdifferential of $F$ at $x$ can be expressed as follows.
\begin{equation}
\partial_* F(x) = \{ \nabla f_{I^+}(x) + \sum_{i \in I^0} \lambda_i \nabla f_i(x) , 0 \leq \lambda_i \leq 1\}.
\end{equation}
The generalized gradient can thus be expressed as a solution to the following quadratic programming (QP) problem, which admits a unique solution in $\lambda$ when all $c_i$ are in generic position -- Sec. \ref{sec:alg-pro}:	
\begin{equation}
\label{eq:qp_clarke}
\nabla_* F(x) = \underset{0 \leq \lambda_i \leq 1} \argmin \left \{ \lvert \lvert \nabla f_{I^+}(x) + \sum_{i \in I^0} \lambda_i \nabla f_i(x) \rvert \rvert^2 \right \}.
\end{equation}
Letting $(\alpha_i)_{i \in I^0}$ be the unique solution to this problem,
we let \footnote{Not to be confused with the sets $I^+(\calC), I^0(\calC), I^-(\calC)$, defined from the sign of the power distance.}
\begin{equation}
\label{eq:IzIpIm}
\begin{cases}
I^0_*(x) & := \{ i \in  I^0(x), 0 < \alpha_i < 1 \}\\
I^+_*(x) & := I^+(x) \cup \{ i \in I^0(x), \alpha_i = 1 \} \\
I^-_*(x) & := I^-(x) \cup \{ i \in I^0(x), \alpha_i = 0 \}.
\end{cases}
\end{equation}
Moreover, we let $\calC^*(x)$ be the cell with three-set partition $I^+_*(x), I^0_*(x),I^-_*(x)$.

\paramini{\textbf{Describing the semiflow of $F$}}

The generalized gradient shines in the following proposition.
Even though $\nabla_*F$ might not be continuous, by convexity of $F$ from any starting point $x$ there exists (see for instance \cite{subgradient_talweg, evolution_problems}) a trajectory $t \mapsto x(t)$ (with $x(0) = x$) called a \textit{semiflow} verifying almost-everywhere (for $t \in \mathbb{R}^+$):
\begin{equation}
x'(t) = -\nabla_* F(x(t)).
\end{equation} 
In particular $F(x(t))$ decreases at rate $\vvnorm{\nabla F_*(x(t))}^2$, and by strong convexity $x(t)$ reaches the argmin of $F$ over $\Rd$ in a finite time, where it is stationary \cite{evolution_problems, gradient_descent_o_minimal}. Given the structure of our $F$, there are three possible behaviors for the semiflow with starting point $x$:
\begin{itemize}
\item  If $x$ is in a full dimensional cell, the semiflow starting from $x$ begins by a segment heading towards $c_{I^+(x)}$,  until it reaches a new cell 
or $c_{I^+(x)}$ which is the minimum.
\item If $x$ lies in a cell of positive codimension and $I^0_*(x)$ is empty, the semiflow enters the non-empty, full dimensional cell $\calC^*(x)$ and follows a straight line in this cell until it meets a new cell, as described as above;   
\item Else, $I^0_*(x)$ is not empty. One can show that if the Clarke QP (Eq. \ref{eq:qp_clarke}) lies in what we call a non-degenerate position\footnote{We say that the QP problem of minimizing $\vvnorm{u + \sum_i \lambda_i v_i}, 0 \leq \lambda_i \leq 1$ lies in a non-degenerate position when the argmin $w$ is such that the set of $i$ such that $\scal{w}{v_i} = 0$ is exactly the set of $i$ such that the coefficient of $v_i$ in the decomposition of $w$ is neither 0 or 1. This condition is standard in the sense that for a given box, for almost all isometries acting on that box, the image box lies in a non-degenerate position.} if and only if  for small $t$ the semiflow enters the non-empty cell of positive codimension $\calC^*(x)$, which is a subset of $S_{I^0_*(x)}$. Points $x$ with a degenerate QP problem are of measure zero, however points $x$ such that the trajectory $x(t)$ reaches a degenerate QP position, making the semiflow intractable, are not. 
\end{itemize}
\subsection{Exact algorithm}  

We develop algorithm \algoexact mimicking the semiflow trajectory
except for the third type of trajectory described above to seek for
the minimum of $F$.  See Algo. \ref{alg:exact} for the pseudo-code -- Sec. \ref{sec:alg-pro}.

It can be decomposed into three so-called main \textit{procedures}, which are \algtele, \algline, \algsphere. The latter is further described using two procedures \algsphereinter and \algminsphereinter. A sixth procedure used in \algline and \algsphere is \algqp. Except for the latter which consists in solving a classical QP programming problem, their pseudo-code can be found in -- Sec \ref{sec:alg-pro}. Procedures \algline, \algsphere are illustrated in Fig. \ref{fig:trajectory}.

(i) If $x_n$ lies in a full dimensional cell $\calC$ (usually at the start of the algorithm), we check if $I^+(\calC)$ contains $\optcell$. If so, we let $x_{n+1}$ be $\optcell$ and we stop the algorithm. Since this step follows property specific to $F$ and not the insight of the semiflow, we call it the \algtele procedure.
 If there is no teleportation, we obtain $x_{n+1}$ from the \algline procedure within $\calC$, which is described as follows. We seek the first point on the half-line starting from $x_n$ heading towards $c_{I^{+}(\calC)}$ meeting another cell, and we let $x_{n+1}$ be this point. This is done by solving for quadratic equations (in $t$) of the form $\vvnorm{x_n + tu - c_i}^2 = R_i^2$.
 
(ii) Else $x_n$ starts in cell of positive codimension. Compute the generalized gradient of $f$ at $x$ as well as the associated $I^+_*(x_n), I^0_*(x_n), I^-_*(x_n)$ with the \algqp procedure (\ie solving Eq. \ref{eq:qp_clarke}). 
\begin{itemize}
\item If the generalized gradient is zero, the minimum has been reached and we can stop the algorithm.
\item If $I^0_*(x_n)$ is empty, follow the \algline procedure described earlier within the full dimensional cell $\calC*(x_n)$. Take $x_{n+1}$ to be the point given by this procedure.
\item Else, the semiflow starting from $x_n$ stays in $S_{I^0_*(x_n)}$ and we follow the \algsphere procedure, which consists in the following. Compute the point $y$ where $f_{I^+_*(x_n)}$ restricted to $S_{I^0_*(x_n)}$ reaches its minimum via a procedure called \algminsphereinter described in more details in the appendix. 
If $y$ is in the cell $\calC^*$, let $x_{n+1}$ be $y$. Else, compute the center $c_S$ and radius $R_S$ of $S_{I^0_*(x_n)}$ via the \algsphereinter procedure. Via the parameterization $[0,1] \mapsto c_S + R_S\frac{(1 -\lambda) x_n + \lambda y - c_S}{\vvnorm{(1 -\lambda) x_n + \lambda y - c_S}}$ of the geodesic on $S_{I^0_*(x_n)}$, check the first point on the geodesic leaving the cell $C^*$. Let $x_{n+1}$ be this point.
\end{itemize}

\ifTWOCOLUMNS

\begin{figure}[htb]
\centerline{\includegraphics[width=.95\linewidth]{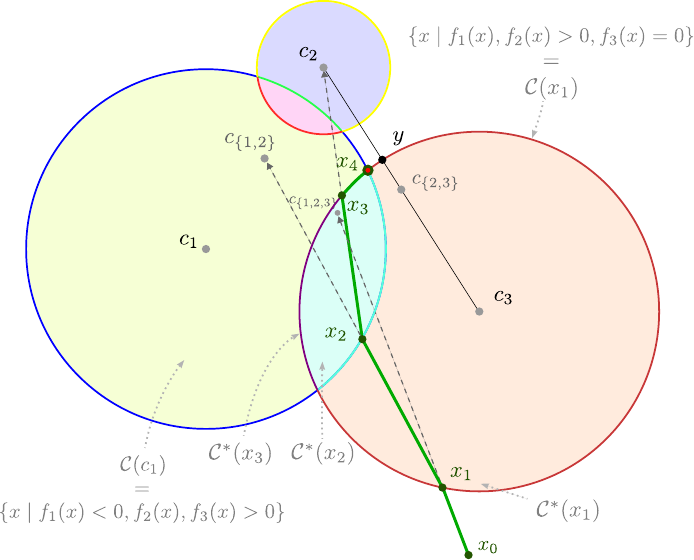}}
\caption{{\bf \algline (from $x_0, x_1, x_2$) and \algsphere (from $x_3$) steps. Underlying trajectories are depicted in dark green. Point $y$ is obtained by \algminsphereinter  point $x_3$.}}
\label{fig:trajectory} 
\end{figure} 

\else

\begin{figure}[htb]
\centerline{\includegraphics[width=.6\linewidth]{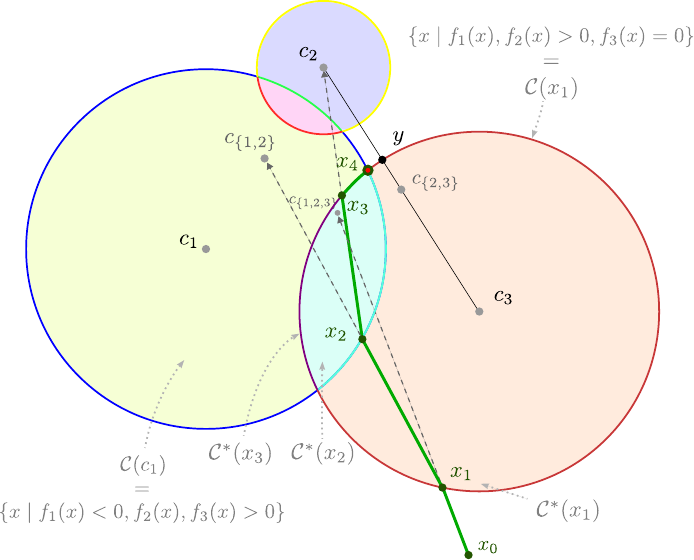}}
\caption{{\bf \algline (from $x_0, x_1, x_2$) and \algsphere (from $x_3$) steps. Underlying trajectories are depicted in dark green. Point $y$ is obtained by \algminsphereinter  point $x_3$.}}
\label{fig:trajectory} 
\end{figure} 

\fi

Following the semiflow ensures that our algorithm converges in a known
number of steps in a certain neighborhood of the point $x^*$ where $F$
reaches its minimum. The number of steps is related to the number of
facets of the Clarke gradient $\partial_* F(x^*)$, which is $3^{c}$
where $c$ is the number of spheres on which $x^*$ lies.
\begin{theorem}[Algorithm convergence]
There exists a neighborhood of the point $x^*$ where $F$ reaches its
minimum, such that for any starting point in this neighborhood the
algorithm converges in at most $3^{c}$ steps, where $c$ is the number
of spheres on which $x^*$ lies.
\end{theorem}

\subsection{Combinatorial complexity}
\label{sub:complexity}
The complexity (number of cells of all dimensions) of the arrangement of $n$ spheres in
$\Rd$ is $O(n^d)$~\cite{toth2017handbook} and the bound is tight in the
worst case. Despite this, we may expect a number of steps polynomial
in $n$ if calculations remain {\em local} in the arrangement, a fact substantiated by our experiments. Note that for small values of $\eta$, the center of mass
provides a warm start to the algorithm.  Indeed the solution of the
minimization problem of $F_{\eta}$ varies continuously in $\eta$, and
the barycenter is solution to the problem with $\eta =0$.

Computations at each step are linear in the total number of
points since \algline (resp. \algsphere) computes the first sphere 
crossed by a line (resp. a sphere geodesics).  Other computations
involved at each step are at most cubic in the number of spheres on
which the current point lies, be it by solving the \algqp problem,
inverting a matrix in \algsphereinter or computing
an affine projection in  \algminsphereinter.
Practically, our algorithm's complexity is nicely behaved--see Sec. \ref{sub:comp_practice}.
\toblack

\subsection{Numerics}
\label{sec:numerics}

\paramini{Arithmetics and number types.}
The algorithm from Sec. \ref{sec:optim} is described assuming the real
RAM model computing exactly with real numbers.  On the other hand,
geometric calculations (predicates, constructions) are known to be
plagued with rounding errors~\cite{kettner2008classroom}.
Serious difficulties may be faced for 
cascaded constructions, which iteratively embed new geometric objects
(the points of the pseudo-gradient trajectory in our case).
Advanced number types combining multiprecision and interval
arithmetics can be used to maintain accurate representations in such
cases. See \eg random walk inside polytopes
\cite{chevallier2022improved} or trajectories of the flow complex (the
Morse-Smale diagram of the distance function to a finite point
set)~\cite{cazals2021frechet}.

In the sequel, we review the numerically demanding operations required by our algorithm,
and refer the reader to Sec. \ref{sec:experiments} for experiments with our python based implementation.

\paramini{Exact solver: predicates and constructions.} 

\sbulem{\bf Solving the \algqp procedure.} The library {\em cvxpy} uses non-exact solvers to find the minimum of a QP problem. 
While those solvers usually give a result up to machine precision, we found out that they were prone to instability in minimizing functionals of the type $\vvnorm{g + A\lambda}^2$ when $g$ is vector of norm largely greater than both 1 and than that of the columns of $A$, 
with constraints $0  \preceq \lambda  \preceq 1$, in the sense that those solvers would claim the problem to be unfeasible. 
The equivalent problem of minimizing $\vvnorm{\frac{1}{\vvnorm{g}}\left ( g + A\lambda \right)}^2$ was sufficient in addressing those issues. 
The precise computation of the vector $\lambda$ is not needed as we only need to check for the index $i$ with respectively $\lambda_i \in \{0 \}, (0,1)$ and $\{1\}$. The entries with values $0$ and $1$ are usually reached with precision greater than machine precision. 

\sbulem{\bf Solving the \algline and \algsphere procedures.} Starting from a point $x$, with a prescribed direction $u$, procedure \algline seeks the first  $t$ such $x +tu$ changes cell, that is, the smallest positive $t$ verifying $\vvnorm{x + tu - c_i}^2 = R^2_i$. This is obtained as roots of a second-degree polynomial. To weaken imprecision we chose to solve this equation with a renormalized vector $u$ of norm 1.  Similarly, given two points $x, y$ on a sphere of radius $R_S$ and center $c_S$, procedure \algsphere computes the first point on the geodesic between $x$ and $y$ changing cell by solving for a quadratic equation. 

\sbulem{Solving the \algsphereinter procedure.} The pair center/radius $c_S, R_S$ used above is obtained as the center and radius of an intersection of spheres. As described in the appendix, the center is obtained through the computations of a projection onto an hyperplane defined by linear equations involving the centers of said spheres. Computing $R_i$ is done by solving for a quadratic equation.

\sbulem{Solving for the \algminsphereinter procedure.} Given an intersection of spheres $S_I$, with $I \subset \{ 1, \dots, n\}, \# I \leq d$ the \algminsphereinter procedure minimizes a function of the form $(f_J)_{|S_I}$ by computing a similar projection on a convex hull.

The genericity assumptions and the robustness of our routines are further discussed
in the SI Section  \ref{sec:genericity}.

\paramini{The  BFGS solver.}
The Broyden–Fletcher–Goldfarb–Shanno quasi-Newton method (BFGS) is
designed to be very efficient on twice differentiable function by
approximating the Hessian matrix without any matrix inversion (in
opposition to Newton's methods), using the gradient. When the gradient
is not given, it is estimated using finite differences.
While the  objective function $\Feta$ is not differentiable,
it is also  known that BFGS  works well in practice for
non-differentiable functions \cite{lewis2012nonsmooth}.
The next section challenges this observation for $\Feta$.

\ifVERSIONCONF
\else
\SCgenericity
\fi



\begin{figure*}[htbp]
\centering
\begin{tabular}{cc}
\includegraphics[width=0.35\textwidth]{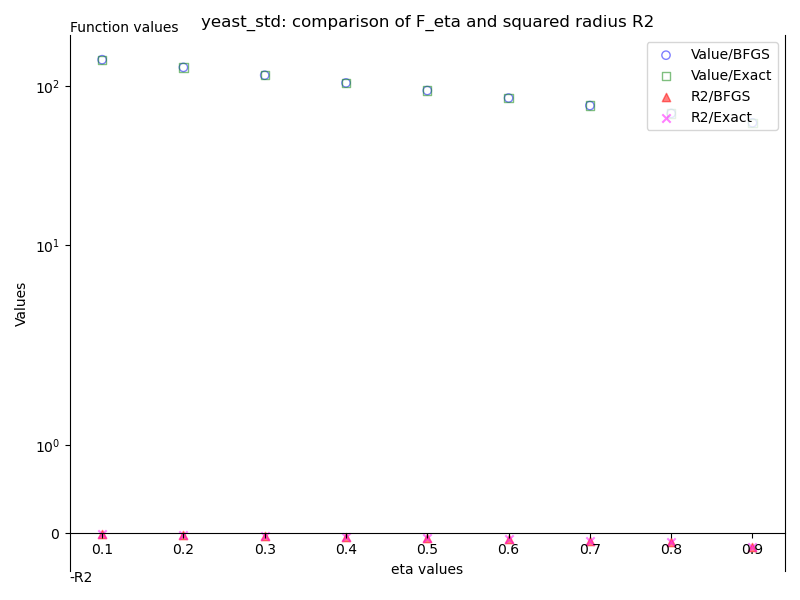} & \includegraphics[width=0.45\textwidth]{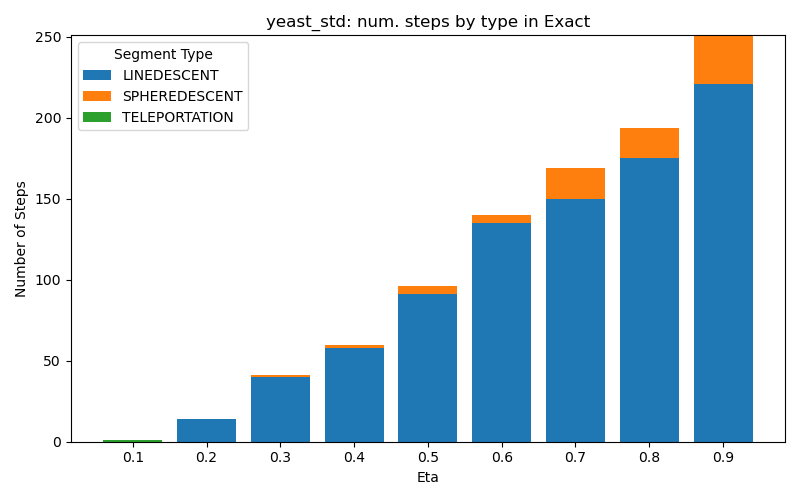}\\
{\scriptsize {\bf (A)} $\Feta$ and $R2$ function of $\eta$} & {\scriptsize {\bf (B)} Step types in trajectory of \algoexact}\\
\includegraphics[width=0.35\textwidth]{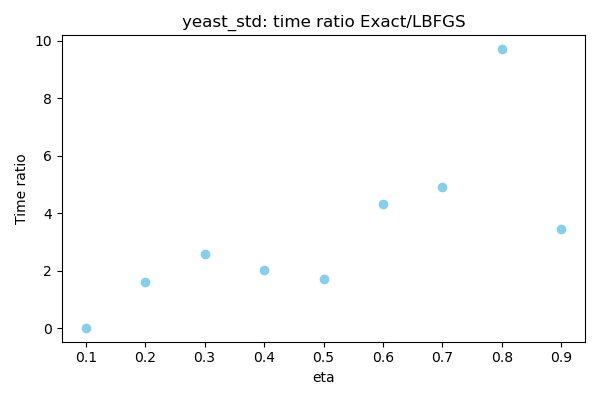} & \includegraphics[width=0.35\textwidth]{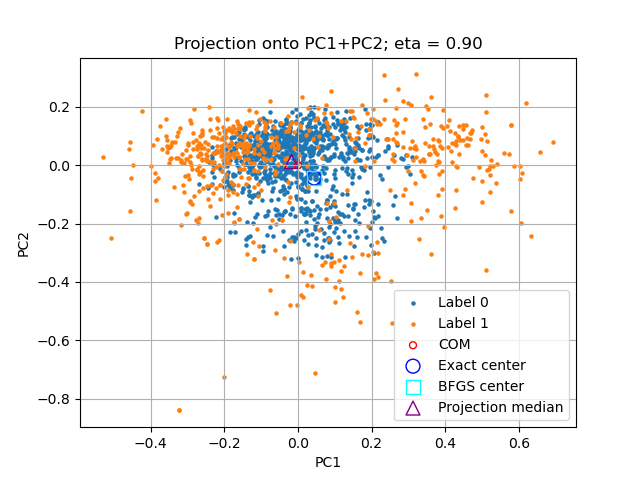}\\
{\scriptsize {\bf (C)} Ratio $\timeexact / \timebfgs$} & {\scriptsize {\bf (D)} Projection plot, $\eta=0.9$}
\end{tabular}
\caption{{\bf Yeast landsat.} This dataset features  $6435$ points in dimension $d=9$.
}
\label{fig:landsat}
\end{figure*}

\section{Spherical clusters: experiments}
\label{sec:experiments}

\subsection{Implementation}

Our implementation of the algorithm from Sec. \ref{sec:optim} using
python and numpy is denoted \algoexact and is termed the {\em exact
  method}. 
It is available from the Core tier of the 
\sblwebhref,  in the
\href{https://sbl.inria.fr/doc/Cluster_spherical-user-manual.html}{Cluster spherical} package.
We compare the solution
yielded by \algoexact against that yielded by \algobfgs -- the
optimization being done with BFGS.  The cluster centers are denoted
$\optexact$ and $\optbfgs$ respectively.

The complete package (python code, documentation, test suite) is
integrated to a fully fledged library containing more than 100
packages, and will be released with the paper.

Calculations were run on a DELL precision 5480 equipped with 20 CPUs
of type Intel(R) Core(TM) i9-13900H, 32Go or RAM, and running FedoraCore~42.
\ifTWOCOLUMNS
\begin{figure}[H]
\centerline{\includegraphics[width=1.06\linewidth]{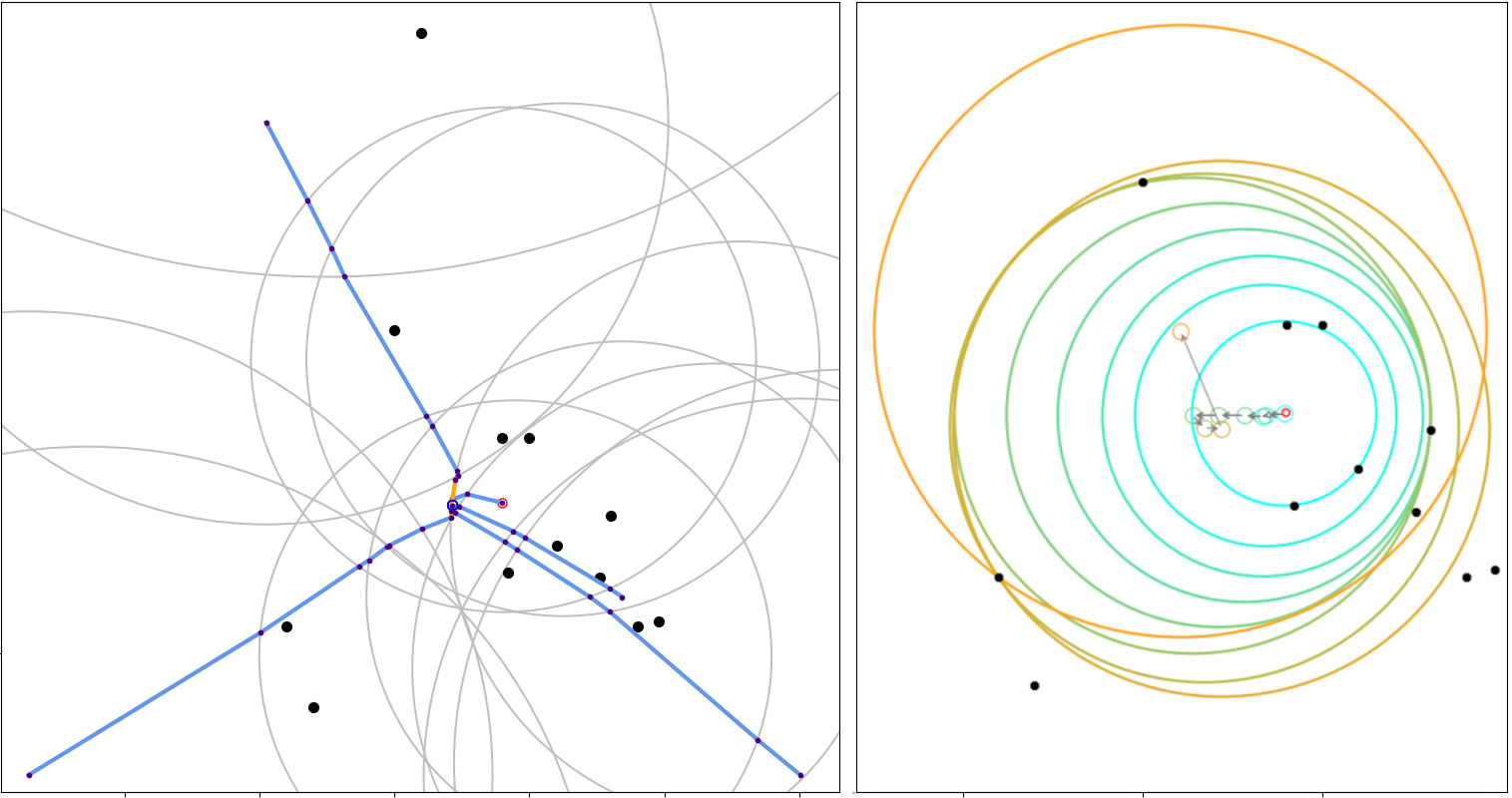}}
\caption{
{\bf Spherical cluster: illustrations on a toy 2D dataset.}
{\bf (Left)} Trajectories from five different starting points, with $\eta = 0.5$ (Line/Sphere descents in blue/orange). 
{\bf (Right)} Evolution of the cluster center for $\eta$ in $[0.1, 0.9]$ be step of 0.1.} 
\label{fig:2D-illustrations} 
\end{figure} 
\else
\begin{figure}[H]
\centerline{\includegraphics[width=.75\linewidth]{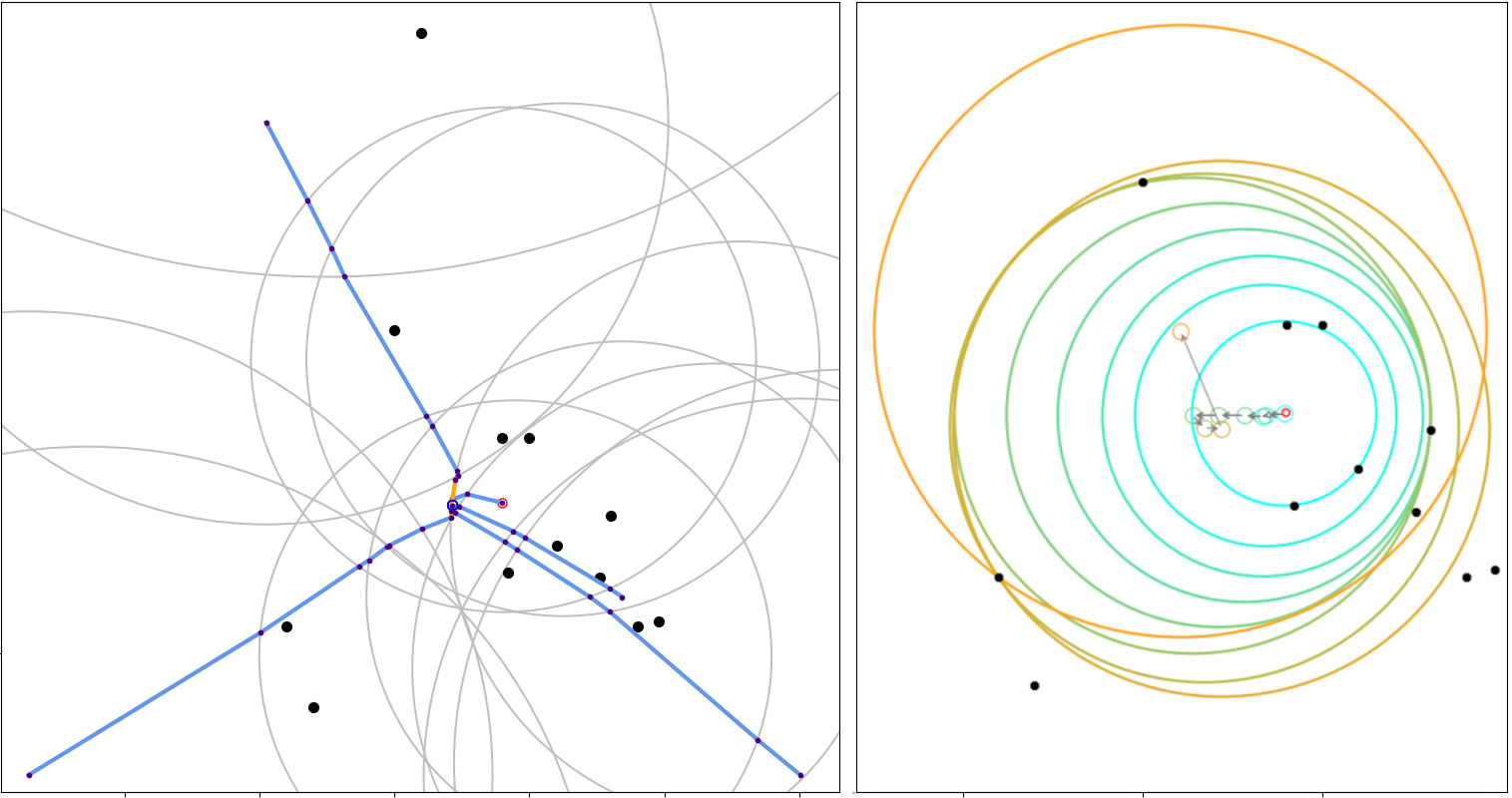}}
\caption{
{\bf Spherical cluster: illustrations on a toy 2D dataset.}
{\bf (Left)} Trajectories from five different starting points, with $\eta = 0.5$ (Line/Sphere descents in blue/orange). 
{\bf (Right)} Evolution of the cluster center for $\eta$ in $[0.1, 0.9]$ be step of 0.1.} 
\label{fig:2D-illustrations} 
\end{figure} 
\fi

\subsection{Contenders, datasets and statistics}

\paramini{Contenders.}
We challenge \algoexact with two contenders denoted \algobfgs and
\algolbfgs respectively, using the BFGS and L-BFGS-B solvers provided
by {\tt scipy.optimize}.  Note that the latter uses an approximation
of the Hessian--as opposed to a $O(d^2)$ sized matrix. 

\paramini{Medium dimensional (MD) datasets. } 
We ran experiments on ten  standards datasets used in clustering
experiments \cite{celebi2013comparative,carriere2025improved}, with
size $n\in [1484,200000]$ and dimension $d\in [9,77]$ -- Table
\ref{tab:datasetCLU}. Following common practice, on a per dataset
basis, we perform a min-max normalization on the coordinates to avoid
overly large ranges.

\paramini{High dimensional (HD) datasets. } 
We use two datasets to explore the effect of high dimensionality.
The  \datasetHMM dataset consists of  $N=1443$ protein sequences whose
biological function is unknown~\cite{vicedomini2022profileview}.  To
identify putative functions, each sequence is scored by $d=400$ Hidden
Markov Models (HMM) corresponding to major known protein functions,
yielding a $d$-dimensional point. Carbone et al. perform hierarchical
clustering on these points (Ward's method), yielding 16 clusters
(sizes in 11..176) of sequences expected to have identical functions.
The 
\href{https://archive.ics.uci.edu/dataset/167/arcene}{Arcene} dataset
contains mass-spectrometric data meant to distinguish 
cancer versus normal patients, and has shape  $(n,d)=(900, 10000)$.
The $d=10000$ features correspond to protein abundances in human sera,
to which distractor features with no predictive power have been added.

\paramini{Parameters.}
For each dataset, we explore values of  $\eta$ in $[0.1,0.9]$ by steps of 0.1 -- nine values in total.

\paramini{Statistics and plots.} We define:\\
\sbulem{SC square radius.}
The square radius with respect to which the power distance is computed, that
is $\eta \hat{\sigma}^2$ -- see Eq. \ref{eq:stdevdist}.

\sbulem{Projection plot.} The plot of all points (data points, center
of mass, SC centers) onto the first two principal directions. Inliers
(resp. outliers) are displayed in blue (resp. orange).
The number of outliers identified by our cluster model,
is denoted $\numoutliersSC$. 
Similarly  $\numoutliersCOM$ stand for the number be outliers defined with respect to a sphere
of the same radius centered at the center of mass.

\sbulem{\bf Dual plot}. Reports $\Feta$ and $R2$ as a function
of $\eta$.  (NB: $R2$ values are represented negated on this plot.)

\sbulem{\bf Stacked barplot.} The plot function of $\eta$ counting the
number of steps of each type (line descent, sphere descent,
teleportation) in \algoexact.

\sbulem{\bf Time ratio plots.} The plots for $\timeexact / \timebfgs$ and
$\timeexact / \timelbfgs$, comparing the running times of 
\algoexact against those of \algobfgs and \algolbfgs
respectively.

\sbulem  {\bf Average outlier cost plots.}
The plots  $\Feta(\optexact)/ \numoutliersSC$ and
$\Feta(\optexact)/ \numoutliersCOM$.

\sbulem{\bf Outlier ratio plot.} The plot 
$\numoutliersCOM/\numoutliersSC$.

\subsection{Spherical cluster model}

\begin{figure}[!htb]
\begin{center}
\begin{tabular}{cc}
\includegraphics[width=.5\linewidth]{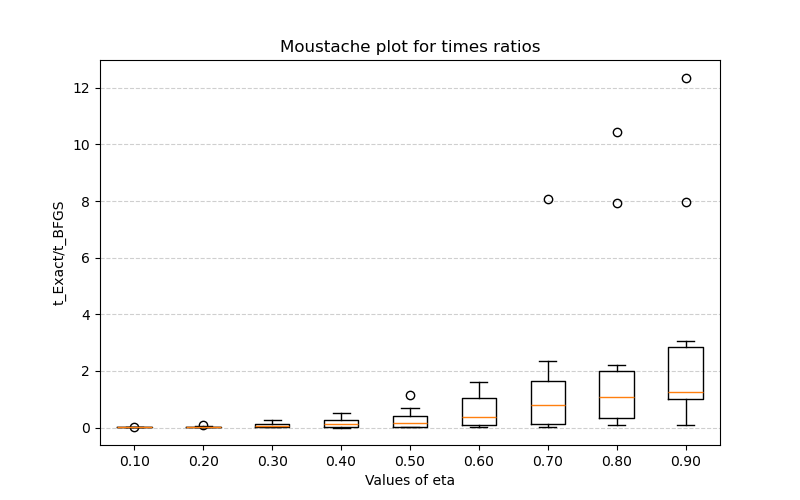}&
\includegraphics[width=.5\linewidth]{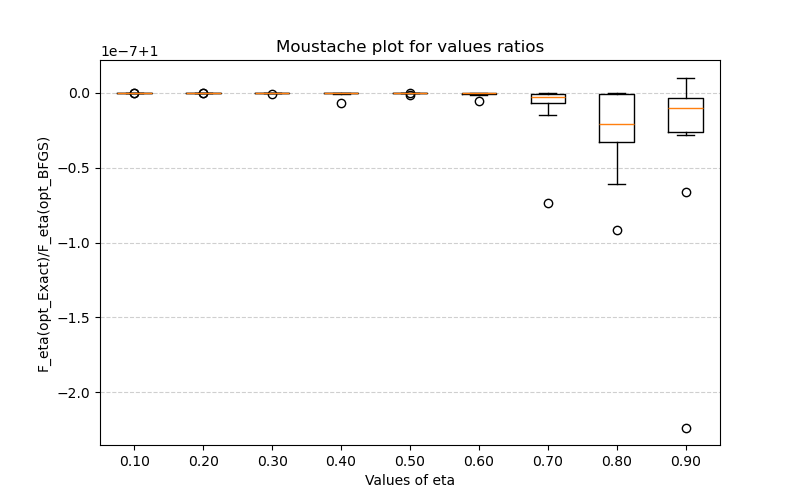}\\
Ratios $\timeexact / \timebfgs$ & Ratios $\Feta{\optexact} / \Feta{\optbfgs}$
\end{tabular}
\end{center}
\caption{{\bf \datasetCLU: \algoexact vs \algobfgs.}}
\label{fig:DatasetCLU-times-values-exact-BFGS}
\end{figure}

\begin{figure}[htb]
\begin{center}
\begin{tabular}{cc}
\includegraphics[width=.5\linewidth]{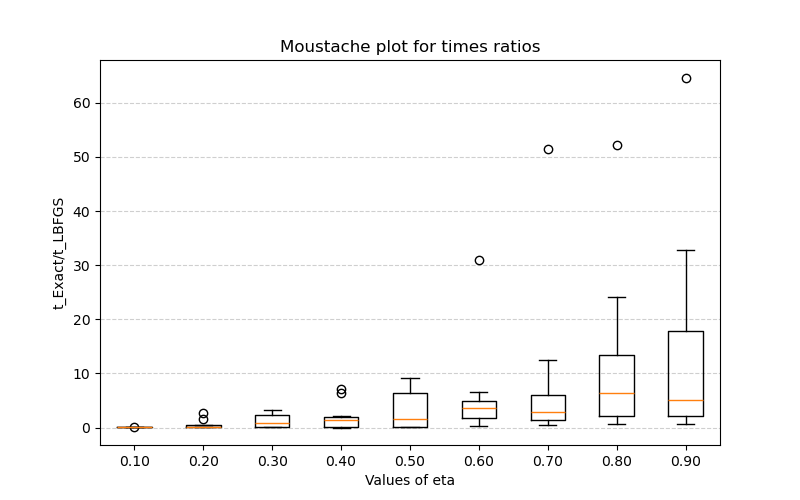} &
\includegraphics[width=.5\linewidth]{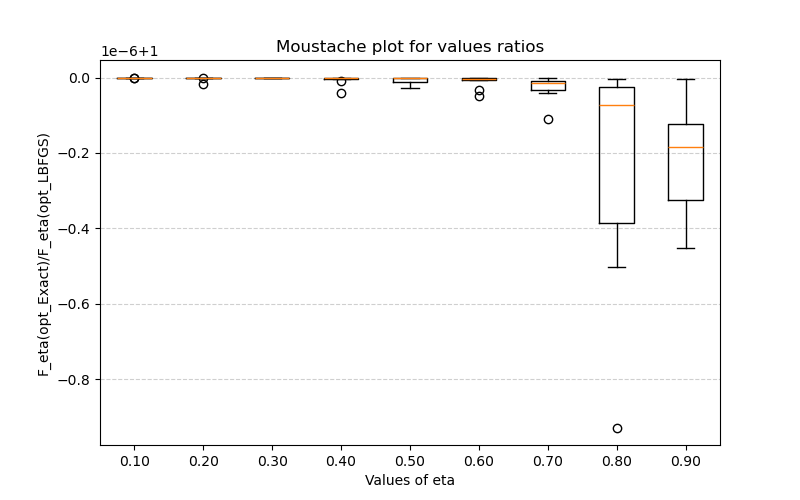}\\
$\timeexact / \timelbfgs$ &  $\Feta{\optexact} / \Feta{\optlbfgs}$
\end{tabular}
\end{center}
\caption{{\bf \datasetCLU: \algoexact vs \algolbfgs.}}
\label{fig:DatasetCLU-times-values-exact-LBFGS}
\end{figure}

\paragraph{Running times and the burden of dimensionality.}
For the dataset \datasetCLU, we first inspect running times using the
ratios $\timeexact / \timelbfgs$ (Fig. \ref{fig:DatasetCLU-times-values-exact-BFGS}, Tab. \ref{tab:DatasetCLU-times-exact-BFGS}),
and $\timeexact / \timelbfgs$  (Fig. \ref{fig:DatasetCLU-times-values-exact-LBFGS}, Tab. \ref{tab:DatasetCLU-times-exact-LBFGS}).
Using median values, the comparison shows that \algoexact is faster than \algobfgs up to up
 to $\eta=0.7$ included, while \algoexact is faster than \algolbfgs up
 to $\eta=0.3$ included. Increasing the value of $\eta$ results in
 larger spheres and more complex arrangements, whence the burden
 observed.

\begin{figure}[htb]
\begin{center}
\begin{tabular}{cc}
\includegraphics[width=.5\linewidth]{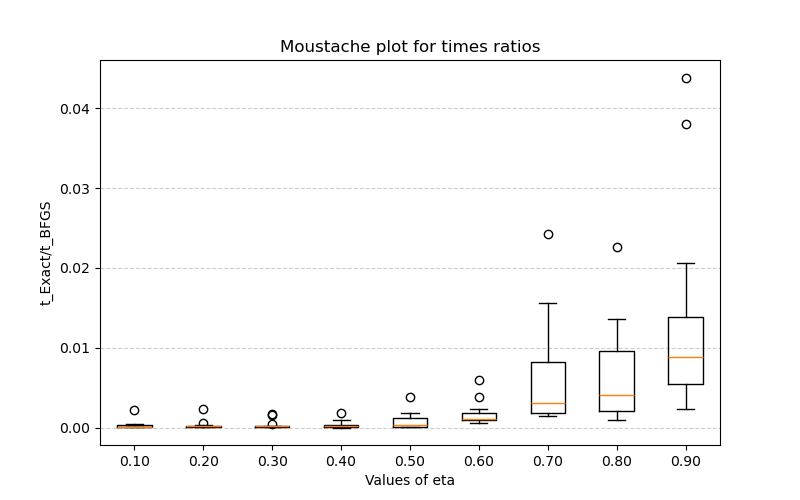}&
\includegraphics[width=.5\linewidth]{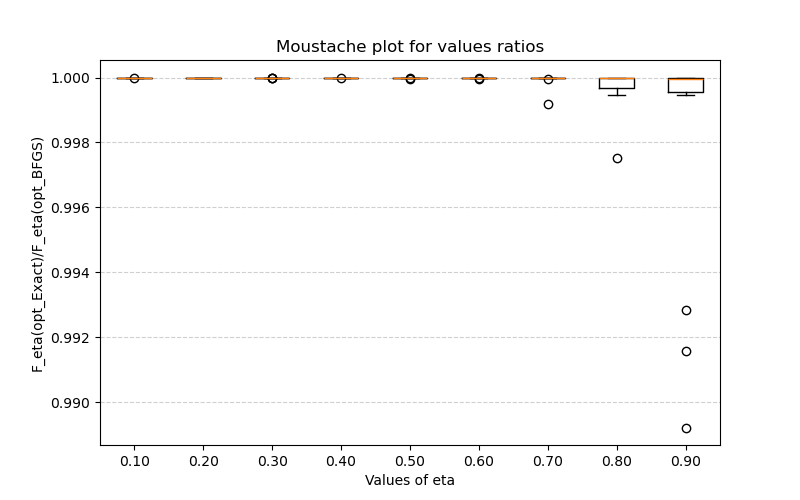}\\
Ratios $\timeexact / \timebfgs$ & Ratios $\Feta{\optexact} / \Feta{\optbfgs}$
\end{tabular}
\end{center}
\caption{{\bf \datasetHMM: \algoexact vs \algobfgs.}}
\label{fig:ProtHMM-times-values-exact-BFGS}
\end{figure}

\begin{figure}[htb]
\begin{center}
\begin{tabular}{cc}
\includegraphics[width=.5\linewidth]{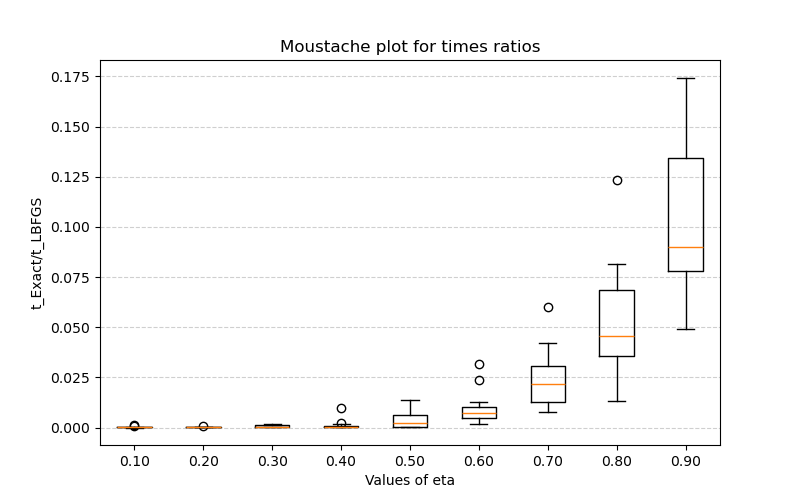}&
\includegraphics[width=.5\linewidth]{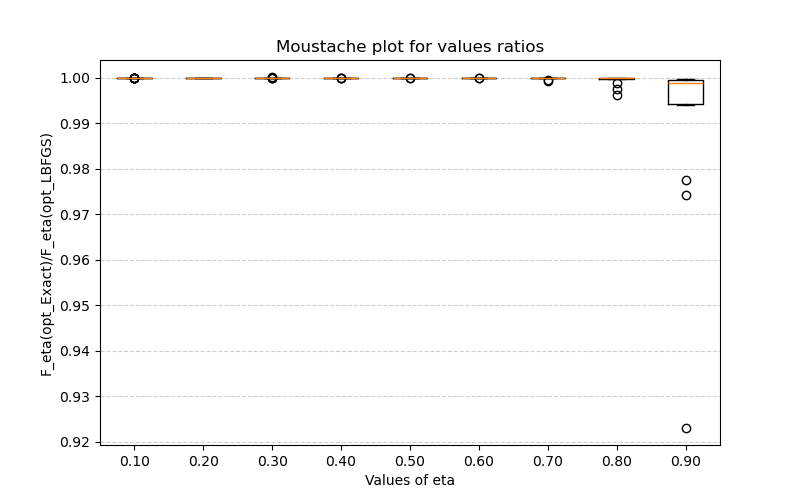}\\
Ratios $\timeexact / \timelbfgs$ & Ratios $\Feta{\optexact} / \Feta{\optlbfgs}$
\end{tabular}
\end{center}
\caption{{\bf \datasetHMM: \algoexact vs \algolbfgs.}}
\label{fig:ProtHMM-times-values-exact-LBFGS}
\end{figure}

We perform the same analysis for the  dataset \datasetHMM,
for 
ratios $\timeexact / \timelbfgs$ (Fig. \ref{fig:ProtHMM-times-values-exact-BFGS}, Tab. \ref{tab:ProtHMM-times-exact-BFGS}),
and $\timeexact / \timelbfgs$  (Fig. \ref{fig:ProtHMM-times-values-exact-LBFGS}, Tab. \ref{tab:ProtHMM-times-exact-LBFGS}).
Using median values again, \algoexact is two for four orders of
magnitude faster than \algobfgs and \algolbfgs.

\begin{figure}[!htb]
\begin{center}
\begin{tabular}{cc}
\includegraphics[width=.5\linewidth]{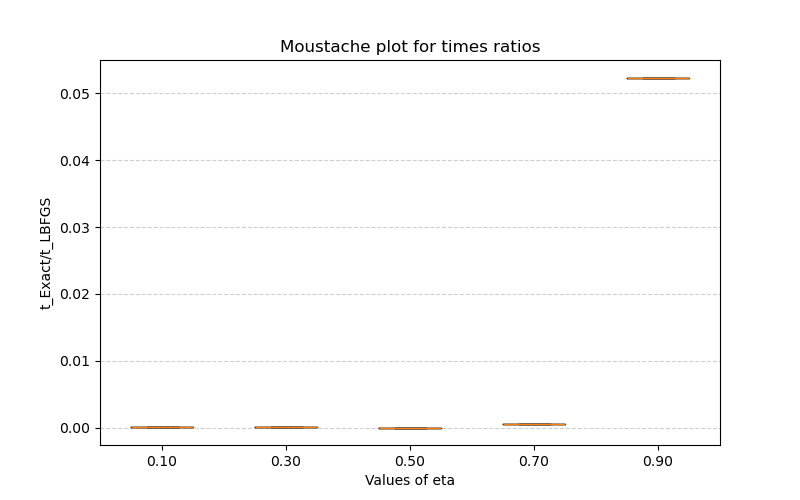}&
\includegraphics[width=.5\linewidth]{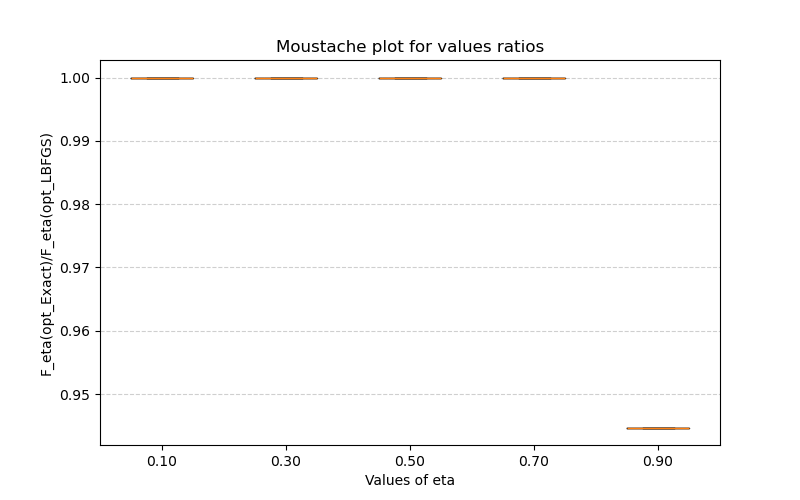}\\
Ratios $\timeexact / \timebfgs$ & Ratios $\Feta{\optexact} / \Feta{\optlbfgs}$
\end{tabular}
\end{center}
\caption{{\bf Arcene: \algoexact vs \algolbfgs.}}
\label{fig:DatasetHD-times-values-exact-LBFGS}
\end{figure}

For the Arcene dataset, BFGS turned out to be unpractical.  We observe
that our exact algorithm is between two and five orders of magnitude
faster than \algolbfgs
(Fig. \ref{fig:DatasetHD-times-values-exact-LBFGS}).

Summarizing, \algoexact is orders of magnitude faster
than \algolbfgs and \algolbfgs for datasets of small/intermediate
dimension and small values of $\eta$, and orders of magnitude faster
than these two methods for high dimensional datasets.

\paragraph{Function values.}
Wen now compare the values yielded by the three contenders:
\algoexact vs \algobfgs: Fig. \ref{fig:ProtHMM-times-values-exact-BFGS} and Table \ref{tab:ProtHMM-values-exact-BFGS};
\algoexact vs \algolbfgs: Fig. \ref{fig:ProtHMM-times-values-exact-LBFGS} and Table \ref{tab:ProtHMM-values-exact-LBFGS}.
While these values are on par for all values of $\eta$, 
we note that the approximate solvers 
are more prone to numerical instabilities,
in particular for \datasetCLU and for \datasetHD.

\paragraph{Outliers and the selection of $\eta$.}
As noticed earlier, the SC center depends both on inliers and outliers.
On all datasets processed, the outlier ratio
$\numoutliersCOM/\numoutliersSC$ lies in the interval $\sim [1, 3]$, which illustrates
the stringency of our  criterion to identify such points.

\begin{figure}[htb]
\begin{center}
\begin{tabular}{cc}
\includegraphics[width=0.404\textwidth]{yeast_std-SC-values-R2-function-of-eta.png} & \includegraphics[width=0.404\textwidth]{yeast_std-SC-stacked-barplot.png}\\
{\scriptsize Dataset yeast\_std: dual plot $\Feta$ and R2} &Dataset yeast\_std: step types in trajectory\\
\includegraphics[width=0.404\textwidth]{yeast_std-SC-time-ratios.png} & \includegraphics[width=0.404\textwidth]{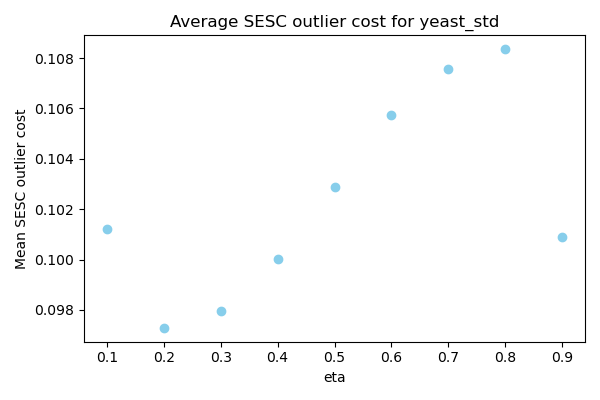}\\
{\scriptsize Ratio $\timeexact / \timebfgs$} &Mean SESC outlier cost\\
\includegraphics[width=0.404\textwidth]{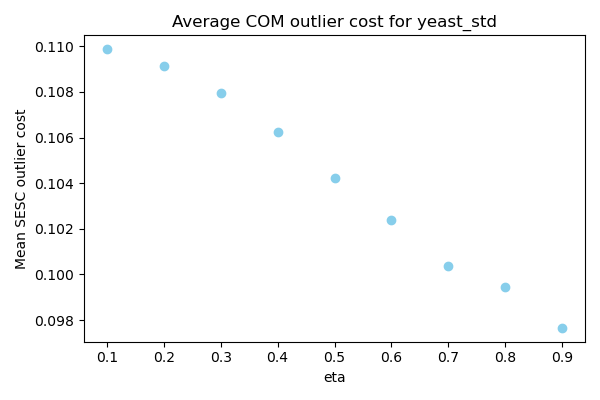} & \includegraphics[width=0.404\textwidth]{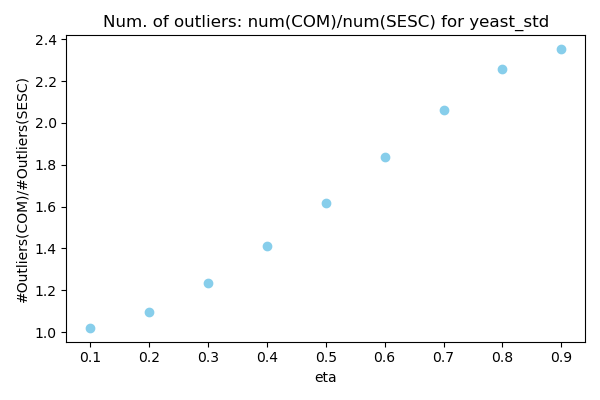}\\
{\scriptsize Mean COM outlier cost} &Outliers: num(COM)/num(SESC)\\
\includegraphics[width=0.404\textwidth]{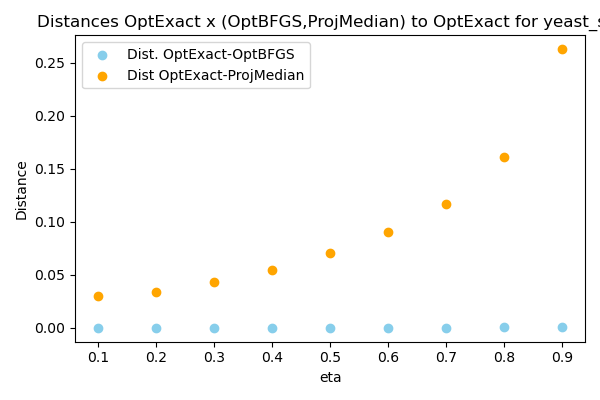}\\
{\scriptsize Distance between points}
\end{tabular}
\end{center}
\caption{{\bf \datasetCLU, dataset yeast\_std -- shape (1484,  9)}}
\label{fig:qq-yeast_std}
\end{figure}

\begin{figure}[htb]
\begin{center}
\begin{tabular}{cc}
\includegraphics[width=0.404\textwidth]{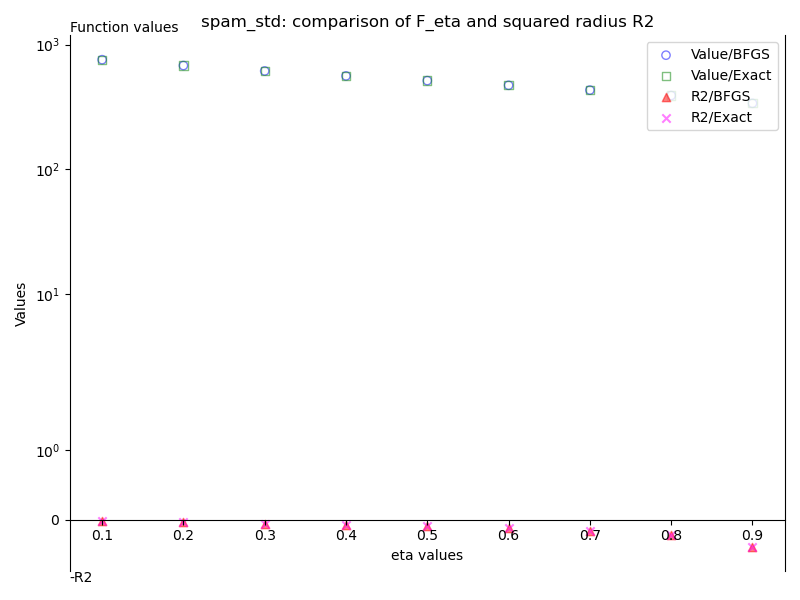} & \includegraphics[width=0.404\textwidth]{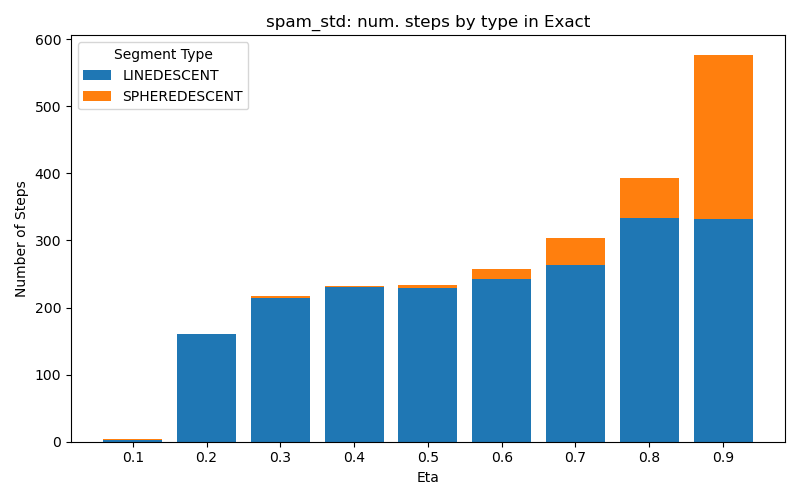}\\
{\scriptsize Dataset spam\_std: dual plot $\Feta$ and R2} &Dataset spam\_std: step types in trajectory\\
\includegraphics[width=0.404\textwidth]{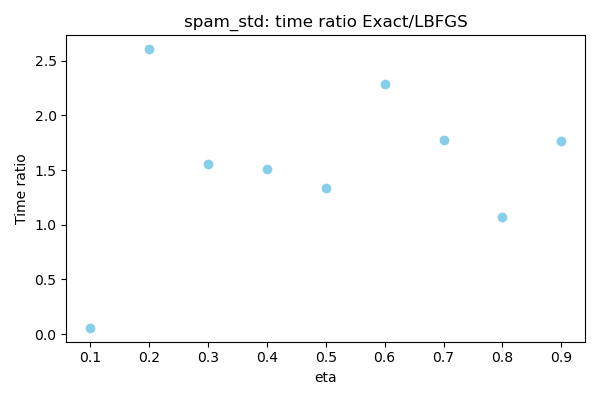} & \includegraphics[width=0.404\textwidth]{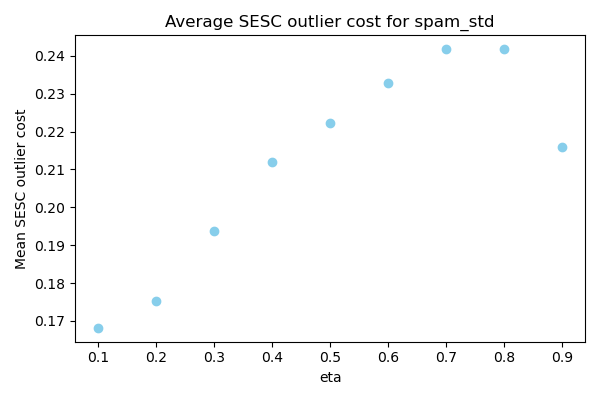}\\
{\scriptsize Ratio $\timeexact / \timebfgs$} &Mean SESC outlier cost\\
\includegraphics[width=0.404\textwidth]{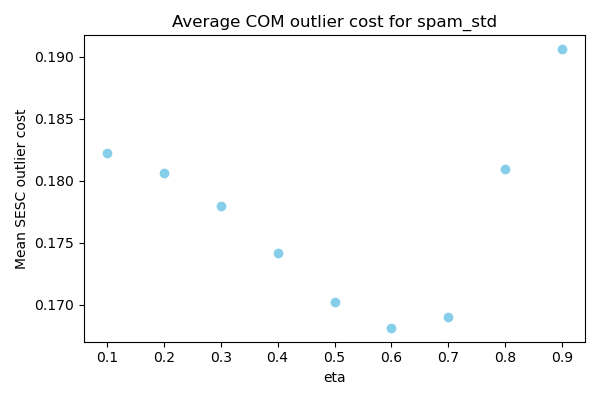} & \includegraphics[width=0.404\textwidth]{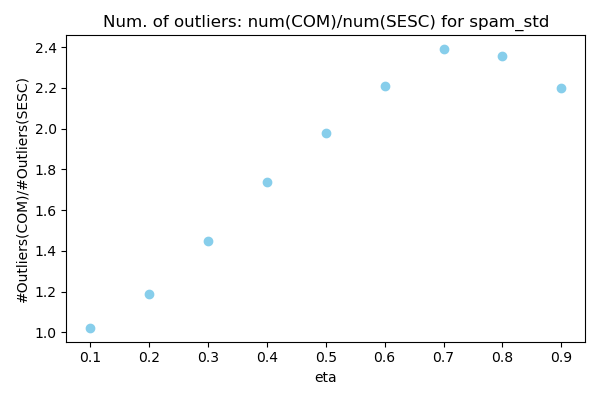}\\
{\scriptsize Mean COM outlier cost} &Outliers: num(COM)/num(SESC)\\
\includegraphics[width=0.404\textwidth]{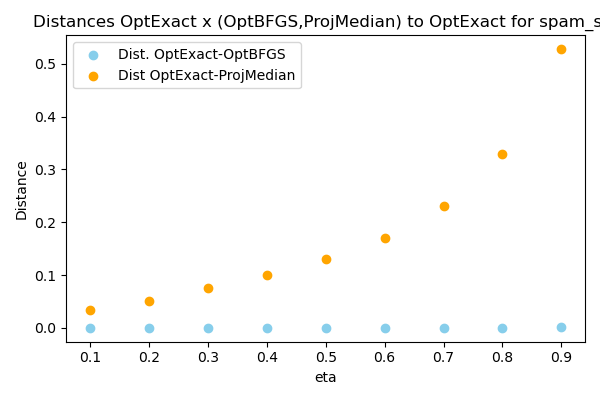}\\
{\scriptsize Distance between points}
\end{tabular}
\end{center}
\caption{{\bf Dataset spam\_std -- shape (4601,  58)}}
\label{fig:qq-spam_std}
\end{figure}

\begin{figure}[htb]
\begin{center}
\begin{tabular}{cc}
\includegraphics[width=0.404\textwidth]{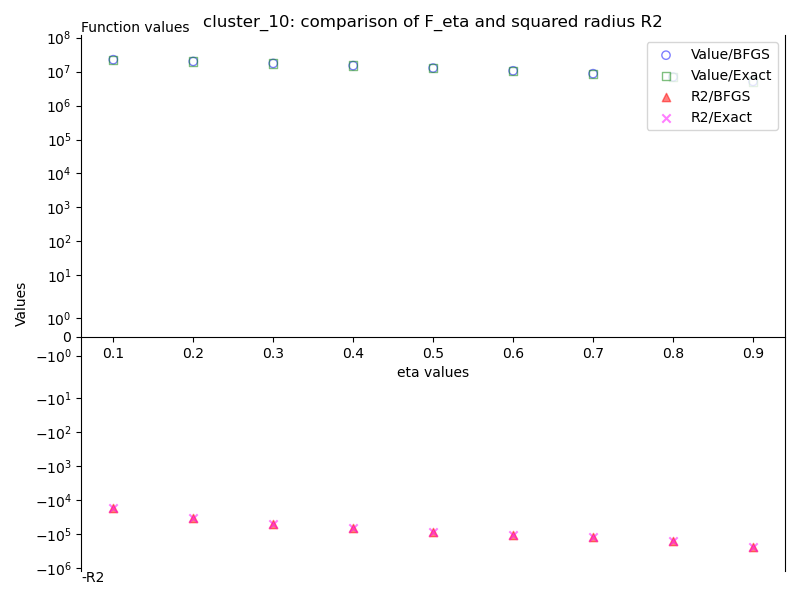} & \includegraphics[width=0.404\textwidth]{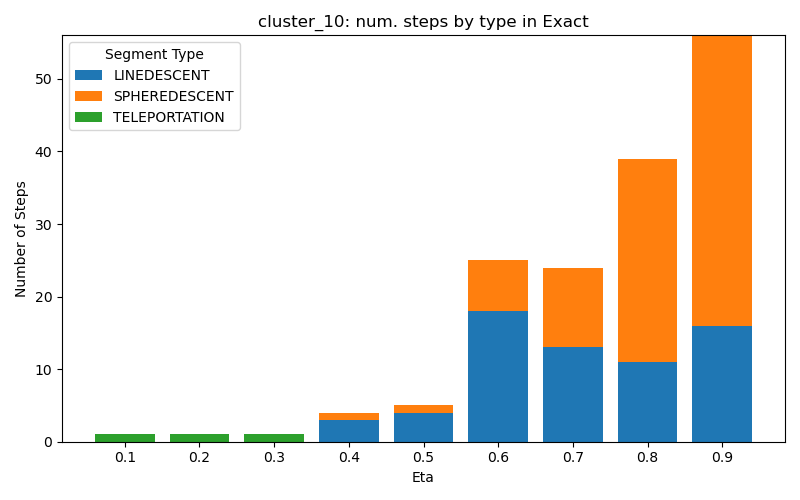}\\
{\scriptsize Dataset cluster\_10: dual plot $\Feta$ and R2} &Dataset cluster\_10: step types in trajectory\\
\includegraphics[width=0.404\textwidth]{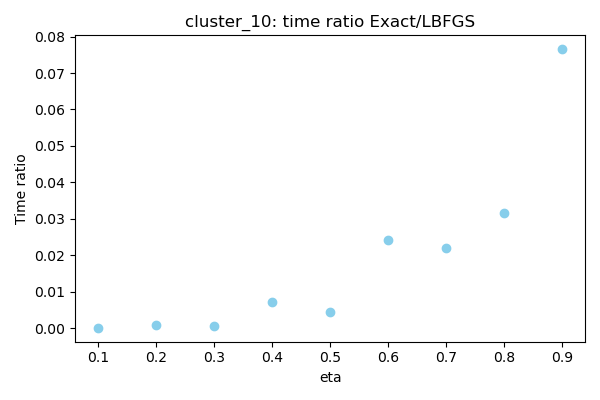} & \includegraphics[width=0.404\textwidth]{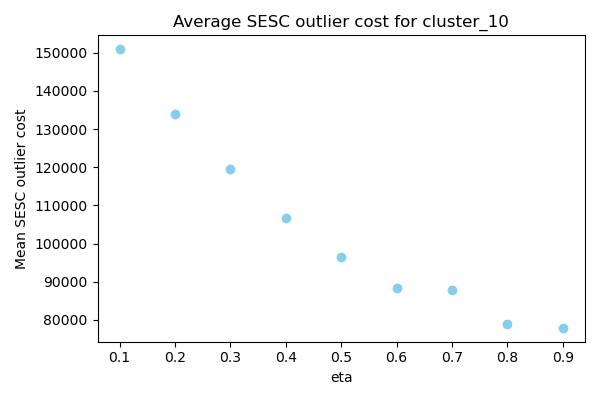}\\
{\scriptsize Ratio $\timeexact / \timebfgs$} &Mean SESC outlier cost\\
\includegraphics[width=0.404\textwidth]{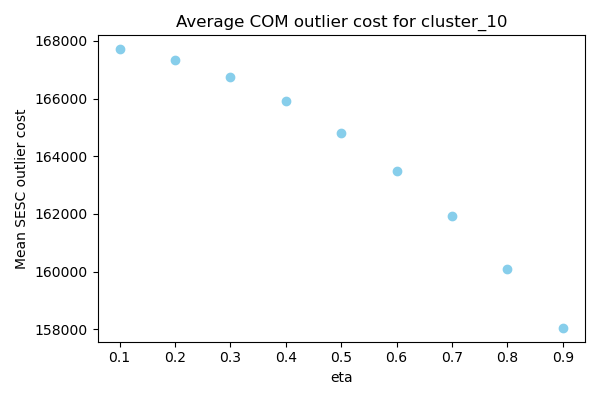} & \includegraphics[width=0.404\textwidth]{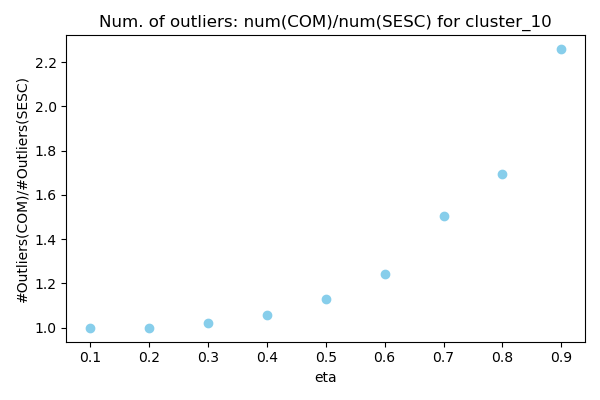}\\
{\scriptsize Mean COM outlier cost} &Outliers: num(COM)/num(SESC)\\
\includegraphics[width=0.404\textwidth]{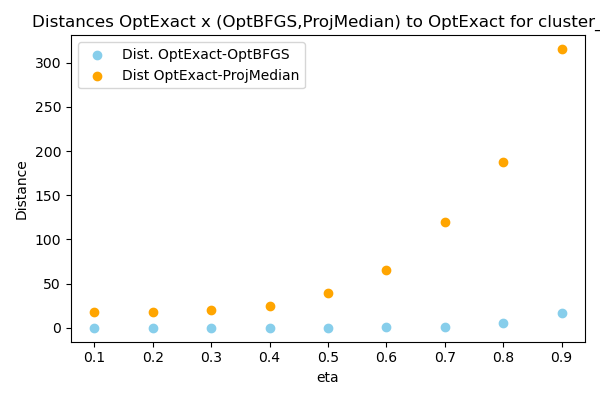}\\
{\scriptsize Distance between points}
\end{tabular}
\end{center}
\caption{{\bf Dataset cluster\_10 -- shape (149,  400)}}
\label{fig:qq-cluster_10}
\end{figure}

\begin{figure}[htb]
\begin{center}
\begin{tabular}{cc}
\includegraphics[width=0.404\textwidth]{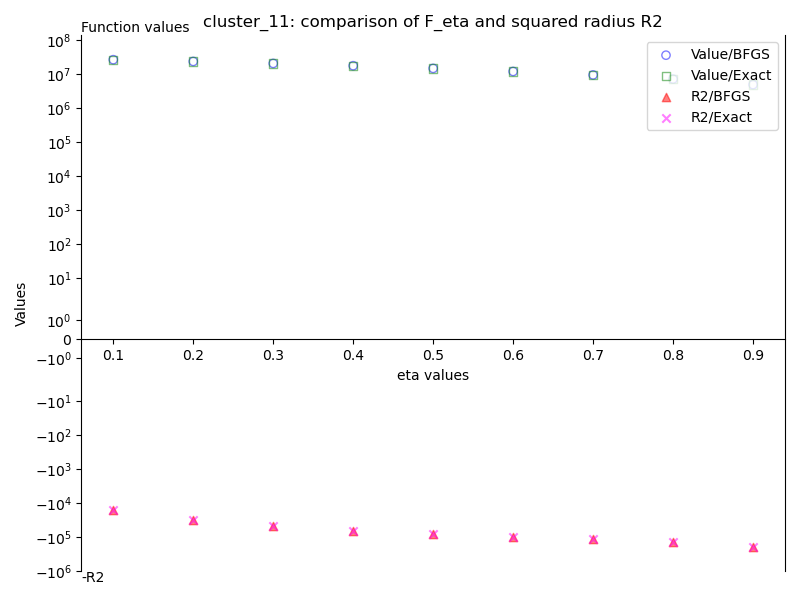} & \includegraphics[width=0.404\textwidth]{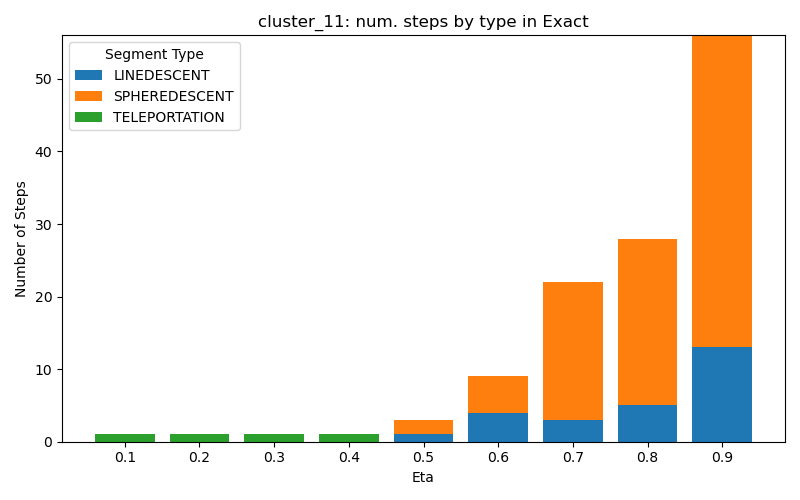}\\
{\scriptsize Dataset cluster\_11: dual plot $\Feta$ and R2} &Dataset cluster\_11: step types in trajectory\\
\includegraphics[width=0.404\textwidth]{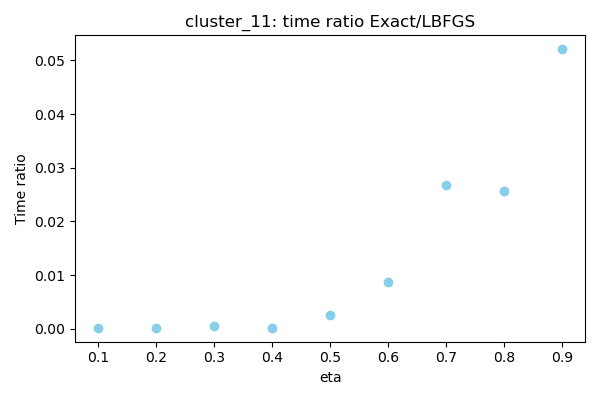} & \includegraphics[width=0.404\textwidth]{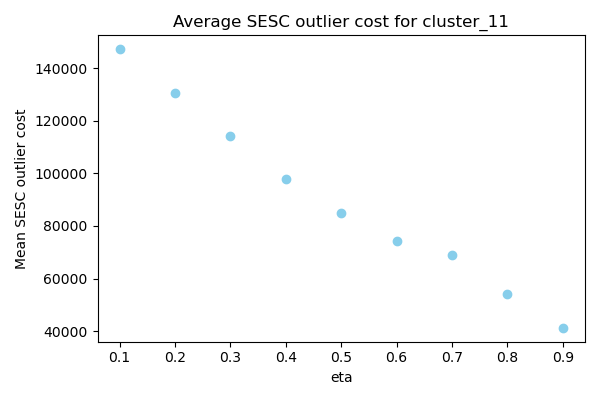}\\
{\scriptsize Ratio $\timeexact / \timebfgs$} &Mean SESC outlier cost\\
\includegraphics[width=0.404\textwidth]{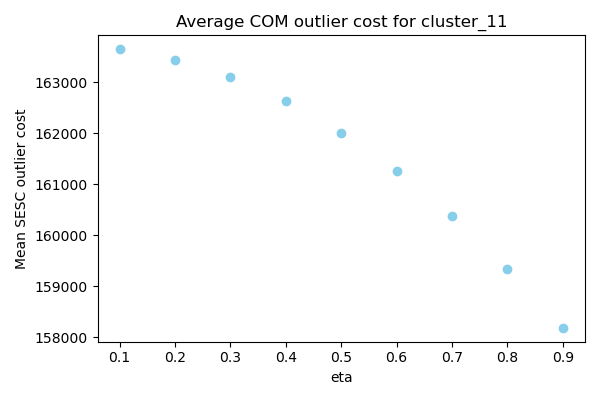} & \includegraphics[width=0.404\textwidth]{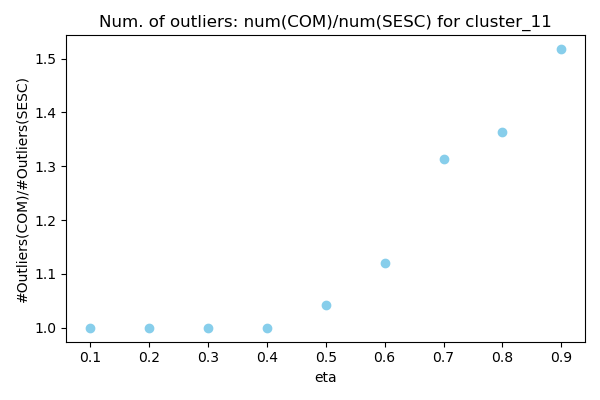}\\
{\scriptsize Mean COM outlier cost} &Outliers: num(COM)/num(SESC)\\
\includegraphics[width=0.404\textwidth]{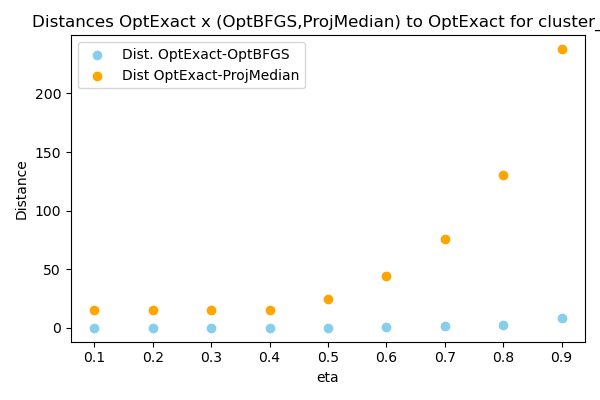}\\
{\scriptsize Distance between points}
\end{tabular}
\end{center}
\caption{{\bf Dataset cluster\_11 -- shape (176,  400)}}
\label{fig:qq-cluster_11}
\end{figure}

The outlier cost plot $\Feta(\optexact)/ \numoutliersSC$ is of particular interest
to capture the scale/cost of outliers.
The general behavior of this plot is a monotonic decrease
(\eg Fig. \ref{fig:qq-cluster_10}, Fig. \ref{fig:qq-cluster_11}), indicating
that {\em capturing} outliers is getting easier when increasing $\eta$.
However several datasets exhibit a non monotonic behavior
(Fig. \ref{fig:qq-yeast_std}, Fig. \ref{fig:qq-spam_std}), showing that {\em gaps} must
be crossed to capture certain outliers.
\toblack

\subsection{Projection median}

We also compare $\optexact$ against the projection median computed
as a weighted average \cite{durocher2017projection}. 
As expected, their distance increases as a function of $\eta$
(Fig. \ref{fig:landsat}(D, Inset), Supporting Information),
showing that the cluster center behaves as a parameterized point set center.
Similar observations hold for the other datasets processed--see SI.

\subsection{Discussion: complexity in practice}
\label{sub:comp_practice}

\paramini{Number of steps and multiplicity of cells.} For all
datasets and whatever the value of eta, we checked that the number of
cells traversed is negligible with respect to the worst case
complexity of the arrangement. We also checked
that in practice the cells visited are so only once, contrary to what
can be found in pathological sphere configurations with bad starting
points. We draw a comparison with the celebrated simplex algorithm,
which has exponential complexity in the worst case, and yet stays in
use after almost 80 years. Our framework is similar in spirit, as the
arrangement is exponential in dimension while the algorithm is
effective in practice. Note that we were unable to build an example
where the number of steps matches the total number of cells. The fact
that our trajectory benefits from a {\em warm start} (warmer as eta
decreases) at the center of mass certainly helps in reducing the
number of cells to be crossed before reaching the minimum, explaining
the efficiency of our algorithm against classical methods (whose
underlying trajectory is not stopped when crossing a cell) when eta is
not close to $1-1/n$.

\paramini{Behavior in high-dimensions.}  In higher dimensions, our
algorithm outperforms the BFGS and L-BFGS. The main factor for this
behavior resides in computation time in between steps cubic in the
number of spheres on which the current point of the trajectory lies
(as discussed in \ref{sub:complexity}), which in practice is low--on
average between 2 and 3 as shown by our experiments: a vast majority
of steps are either \algline procedures or \algsphere on a small
number of spheres (rarely more than five even in high dimensions).
\toblack

\section{Outlook}

Spherical clusters embedded in affine spaces of fixed dimension provide
useful insights into the geometry of high dimensional point clouds.

Our work shows spherical clusters are well defined by a non smooth
strictly convex problem.  We also show that this optimization problem
is well poised and can be solved by an exact iterative procedure
following a semiflow on a stratified complex defined by an arrangement
of spheres. Quite remarkably, BFGS also solves all the instances we
processed to satisfaction.
Yet, the exact solver is orders of magnitude faster than BFGS
based heuristics for high dimensional datasets (say $d>100$), and for
dataset of medium dimensionality and small values of $\eta$.
Our experiments also show that the center of spherical clusters behave
as a high dimensional median parameterized by the fraction $\eta$ of
the variance of distances between the cluster center and all points.

Our work calls for future developments in two directions.
From a theoretical standpoint, understanding the complexity of our
exact method as a function of $\eta$ appears as a challenging problem.
Practically, spherical clusters were proposed to define mixtures
of such clusters embedded into affine spaces of arbitrary codimension.
We anticipate that such models will prove extremely useful in data analysis
at large, providing at the same time compact clusters capturing the intrinsic
dimension of the data.

\noindent{\bf Acknowledgments.} Alessandra Carbone is 
acknowledged for the ProteinHMM dataset.
This work has been supported by the
French government, through the 3IA C\^ote d’Azur Investments
(ANR-23-IACL-0001), and the ANR project Innuendo (ANR-23-CE45-0019).


\beginSI

The supporting information is structured as follows:
\begin{itemize}
\item Section \ref{sec:alg-pro}: Algorithms and proofs
\item Section \ref{sec:exp}: Experiments
\end{itemize}

\section{Algorithms and proofs}
\label{sec:alg-pro}

In this section of the Appendix, we develop the theoretical background and prove the results of the article.
More precisely, 
\begin{itemize}
\item  We prove several intermediate results needed for some propositions used in the article. Intermediate results will be called \textit{lemmas} while the results quoted in the article will be called \textit{propositions}.
\item We precisely describe procedures \algline and \algsphere, as
  well as the procedures \algsphereinter and \algminsphereinter 
  used internally. We also explain the maths behind them.
\item We describe the full \algexact algorithm and prove its convergence.
\end{itemize}

\subsection{Lemmas and propositions}

With the centers $c_i$ and radii $R_i$  as  defined in Eqs. \ref{eq:sink-spec}, we have:
\begin{lemma}[Freeness of the $\nabla f_i(x)$]
\label{lem:free_gradient}
For any $x$ in $R^d$ and for any $J \subset \{ 1, \dots, n \}$, the rank of the family $(\nabla f_i(x))_{i \in J}$  
is the dimension of the affine hull of $x$ and $(c_i)_{i \in J}$.

As a consequence, if the centers $c_i$ are in generic position, for any $x \in \Rd$ the family $(\nabla f_i(x))_{i \in I^0(x)}$ is free.
\end{lemma}

\begin{proof}
The rank of said family is the rank of the vectors $(x - c_i)_{i \in J}$, which is exactly the dimension of the affine hull of $x$ and $(c_i)_{i \in J}$.
For the second point, if $I^0(x)$ is empty (resp. a singleton) the result is trivial by convention (resp. because no radius is zero). Assuming $k = \#I^0(x) \geq 2$, $x$ belongs to the intersection $S_{I^0_(x)} = \bigcap_{i \in I^0(x)} S_i$ which is a $d - k$ dimensional sphere centered around a point $c_S$
of $\mathrm{Aff}((c_i)_{i \in I^0}$ orthogonal to the previous affine space, with radius $R_S$ - see Lemma \ref{lem:sphere_intersection} for the exact computations. As a consequence the dimension of the affine hull of $x$, $(c_i)_{i \in I^0(x)}$ is 1 + the dimension of the affine hull of $(c_i)_{i \in I^0(x)}$ which is $k-1$ by genericity of the centers.  
\end{proof}

The following lemma describes the full-dimensional cells in any neighborhood of a point. It is illustrated in Fig. \ref{fig:neighboring_cells} below.

\begin{lemma}[Full-dimensional cells around a point]
\label{lem:neighboring_cells}
Assume that the family $\{ c_i\} $ lies in a generic position. 
Let $x \in \Rd$ be a point with $\#I^0(x) = k(x) > 0$. Then any neighborhood of $x$ encounters exactly $2^{k(x)}$ cells of full-dimension, and any such cell is determined by $I^+ = I^+(x) \cup A$, $I^- = I^-(x) \cup B$, where $A \cup B$ is a 2-partition of $I^0(x)$.
\end{lemma}

\begin{proof}
It is clear that the full dimensional cells possibly existing around $x$ are included in the one described in the proposition.

 Reciprocally, let $A \subset I^0(x)$ and $B = I^0(x) \setminus A$. Consider the set $U_{A}$ the cone of vectors such that $\scal{\nabla f_i(x)}{u} \geq 0$,$\scal{\nabla f_j(x)}{u} \leq 0$ for all $i \in A$, $j \in B$. This is the polar (or dual) cone to the cone $T_A$ generated by the vectors $(-\nabla f_i(x))_{i \in A}$ and $(\nabla f_j(x))_{j \in B}$; as such $U_A$ has non-empty interior if and only if $T_A \cap (-T_A) = \{ 0 \}$. This condition is satisfied (for every $A \subset I^0(x)$) as the family $(\nabla f_i(x))_{i \in I^0(x)}$ is free by the previous lemma. Any $u$ in the interior of $U_A$ satisfies $\scal{\nabla f_i(x)}{u} > 0$, $\scal{\nabla f_j(x)}{u} < 0$ for any $i \in A, j \in B$ and thus $I^+(x +tu) = I^+(x) \cup A$, $I^-(x +tu) = I^-(x) \cup B$ for $t > 0$ small enough.
\end{proof}

\begin{figure}[htb]
\centerline{\includegraphics[width=.5\linewidth]{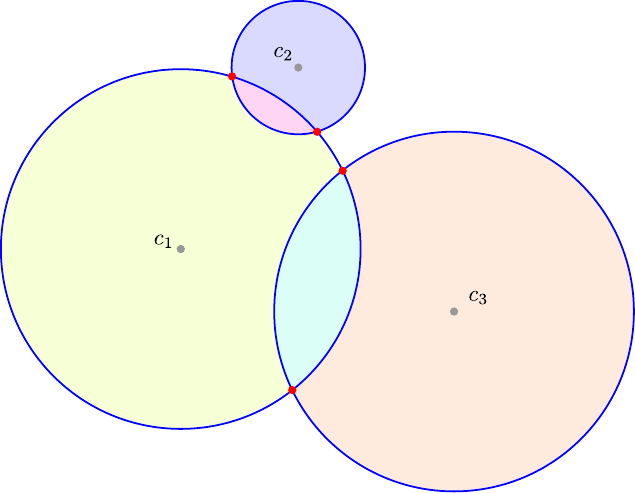}}
\caption{{\bf Neighboring cells.}
Each red dot has $I^0$ of cardinal two and $2^2$ neighboring cells of full dimension. In blue, points with $I^0$ of cardinal 1 have $2^1$ neighboring cells.}
\label{fig:neighboring_cells} 
\end{figure} 

\begin{figure}[htb]
\centerline{\includegraphics[width=.5\linewidth]{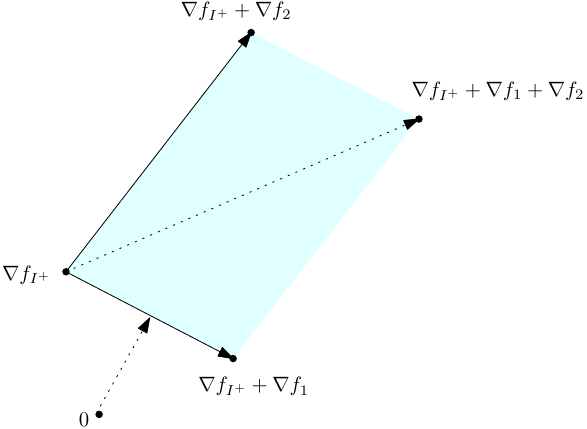}}
\caption{{\bf Clarke gradient.} The Clarke gradient is the convex hull, and the generalized gradient
if the projection of the origin onto that convex hull.
On this example $I^0= \{ 1,2\}$, and $I^0_{*} = \{ 1\}$.
}
\label{fig:clarke-gradient} 
\end{figure}

It is clear that $F$ is smooth at point $x$ if and only if $x$ does not lie on a sphere. The Clarke differential of $F$ at $x$ will allow us to study $F$ in a neighborhood of $x$. Thanks to Prop. \ref{alg:LineDescent} the Clarke gradient of $F$ has an explicit representation (Fig. \ref{fig:clarke-gradient} for an illustration):

\begin{proposition}[Clarke gradient of $F$]
\label{prop:clarke_F}
For any $x \in \Rd$ we have :
\begin{equation}
\label{eq:clarke_F}
\clarke F(x) = \{ \nabla f_{I^+(x)} + \sum_{i \in I^0(x)} \lambda_i \nabla f_i, 0 \leq \lambda_i \leq 1\}.
\end{equation}
When $0 \notin \clarke F(x)$, the minimum of $\lambda \mapsto \vvnorm{\nabla f_{I^+(x)} + \sum_{i \in I^0(x)} \lambda_i \nabla f_i}$ is attained at a unique point.
\end{proposition}

\begin{proof}
Recall that the Clarke gradient of a Lipschitz function is obtained as the convex hull of limits of nearby gradients \cite{clarke1997nonsmooth}.
Denote by $B$ the set on the right-handside of Eq. \ref{eq:clarke_F}. $B$ is a convex parallelepiped whose vertices are $\nabla f_{I^+(x) \cup A}(x) = \nabla f_{I^+(x)}(x) + \nabla f_A(x)$ where $A$ ranges among the subsets of $I^0(x)$ (with the convention that $\nabla f_{\emptyset} = 0$); these are also exactly points determining the convex hull in the definition of $\clarke F(x)$.

Now the gradients $(\nabla f_i)_{i \in I^0(x)}$ are free by Lemma \ref{lem:free_gradient}. It follows that $\lambda \mapsto \vvnorm{\nabla f_{I^+(x)} + \sum_{i \in I^0(x)} \lambda_i \nabla f_i}^2$ is strictly convex and thus admit only one minimizer on the convex set $[0,1]^n$.
\end{proof}

Recall the definition of the sets  $I^0_*(x), I^+_*(x),  I^+_*(x)$ in Section \ref{sec:sesc-opt},
from which one defines the cell $\calC^*(x)$ of the arrangement containing $x$.
We have the following:
\begin{proposition}
When $I^0_*(x)$ coincides with $ \{ i, \scal{\nabla_* F(x)}{\nabla f_i(x)} = 0 \}$, we say that the Clarke QP problem is \emph{non-degenerate} at $x$. In this case the exact gradient flow of $F$ begins by a curve inside $\calC^*(x)$.
\end{proposition}

\begin{proof}
We will show that for $t$ positive and small enough, the semiflow trajectory $t \mapsto x(t)$ starting at $x$ lies in $\calC^*(x)$ by showing separately that
\begin{itemize}
\item $x(t)$ stays in $S_{I^0_*(x)}$ for $t$ small enough and positive;
\item $I^+(x(t))= I^+_*(x)$ and $I^-(x(t)) = I^-_*(x)$ for $t$ small enough.  
\end{itemize}

Remark that the coefficients $\alpha_i$ defining $\nabla F_*(x)$ are obtained through the affine projection of $0$ onto the affine set generated by $\nabla f_{I^+_*(x)} + \nabla f_i(x)$ where $i$ ranges among $I^0_*(x)$ by the non-degenerate condition. This is equivalent to $\lambda_i(x)$ being the solution of the following non-singular linear equation:
\begin{equation}
\sum_{i \in I^0_*(x)} \lambda_i (x) \scal{\nabla f_i (x)}{ \nabla f_j (x)} = - \scal{\nabla f_{I^+_*(x)}}{ \nabla f_j (x)} \qquad \text{ for every } j \in I^0_*(x).
\end{equation}

Equivalently, putting $A$ the full-rank matrix whose rows are $(\nabla f_i(x))_{i \in I^0_*}$, $\lambda$ is a solution to
\[ 
\latrans{A(x)} A(x) \lambda(x) = - \latrans{A(x)} \nabla f_{I^+_*}(x),
\]
and the corresponding generalized Clarke gradient is equal to $\phi(x) = A(x)\lambda(x) + \nabla f_{I^+_*(x)}(x)$. Since $x \mapsto A(x)$ is a smooth map, the ODE $z'(t) = \phi(z(t))$ starting with $z(0) = x$ admits a solution for every $t \in \Rd$. 
This solution coincides with the exact gradient flow for small $t$, as by continuity we still have $0 < \lambda_i(z(t)) < 1$ for small $t$, so that the point of least norm in $\Aff(\nabla f_{I^+_*(z(t))} + \nabla f_i(z(t))_{i \in I^0_*(z(t)) \cup {\emptyset}}$ coincides with the point of least norm within the parallelogram with generated by the same vectors. 
This ensures that $\scal{\nabla f_i(t)}{z'(t)} = 0$ for any $i \in I^0_*(z(t))$ so that the trajectory stays on $S_{I^0_*(x)}$. 
Moreover, for $t$ small enough we have $I^+(z(t)) = I^+(x) \cup {i \in I^0(x), \lambda_i(x) = 1 }$ and $I^-(z(t)) = I^-(x) \cup {i \in I^0(x), \lambda_i(x) = 0 }$, so that the Clarke gradient $\clarke F(z(t))$ is the parallelogram generated by the previous vectors $\nabla f_{I^+_*(x)}(x) + \nabla f_i(x)$, where $i$ ranges among $I^0_*(x)$.
\end{proof}

\subsection{Procedures and their justification}

\algname{\algline}
\begin{algorithm}[H]
\caption{Compute the first point along the semi-line $x+tu$, $t > 0$ changing cell. Compute the cell to which this point belongs.}
\label{alg:LineDescent}
\begin{algorithmic}
\Require{A set of centers $c_i$ and radii $R_i$, an initial point $x$, a directing vector $u$, sets of indices $I^+, I^0, I^-$.}
\Ensure{$z = x +tu$ is the first point along the semi-line $x+tu$, $t > 0$ changing cell. $I^+, I^0, I^-$ are the set of indices at $z$.} 
\State $t \gets + \infty$
\State ChangingIndices $\gets\emptyset$ \Comment{{\togray Set of indices in $I^+ \cup I^-$ to which $z$ belongs - usually a singleton.}}
\For{$i \in I^+ \cup I^-$}
 \State $t_i \gets$  Smallest positive solution to $\vvnorm{x + tu - c_i}^2 = R_i^2$ (with $t_i = \infty$ if no solution).
 \If{$t_i = t$ and $t_i \neq \infty$}
 	\State Add $i$ to ChangingIndices
 \ElsIf{$t_i < t$}
 	\State ChangingIndices $\gets \{ i \}$.
 	\State $t \gets t_i$.
 \EndIf
 \EndFor 
 \State Add elements of ChangingIndices to $I^0$, Remove elements from ChangingIndices off of $I^+$, $I^-$.
\State \Return{$z = x + tu$, $I^0$, $I^+$, $I^-$}
\end{algorithmic}
\end{algorithm}

The \algsphere procedure is less straightforward. It needs two subprocedures : \algsphereinter and \algminsphereinter. The first one is done by solving for a linear system.

\begin{lemma}[Computations in the \algsphereinter procedure.]
\label{lem:sphere_intersection}
Let $2 \leq j \leq d$ and let $S_1, \dots S_j$ be spheres in $\Rd$ with centers $c_1, \dots, c_j$ (in generic position) and radii $R_1, \dots, R_j$ intersecting transversely.
Then $S = \bigcap_{1 \leq i \leq j} S_i$ is manifold of dimension $d-j-1$ described as the intersection of a sphere of radius $R_S$ and center $c_S$ intersected with an affine subspace given by the $j-1$ equations
\begin{equation}
\scal{x}{(c_1 - c_i)} = K_i,
\end{equation}
for all $2 \leq i \leq j$ with $K_i := \frac{1}{2} \left (R_1^2 - \vvnorm{c_1}^2 - R_i^2 + \vvnorm{c_i}^2 \right )$.
Moreover, the center $c_S$ is the only convex combination of $c_1, \dots, c_j$ belonging to this affine subspace, and it is obtained as the solution of the linear system with $j$ equations and $j$ unknowns:
\begin{equation}
\label{eq:centre_sphere}
\left \{
\begin{array}{rlr}
\sum_{i} \lambda_i \scal{c_i}{(c_1 - c_k)} & = K_k & \text{ for every } 2 \leq k \leq j; \\
\sum_{i} \lambda_i & = 1. &
\end{array}
\right.
\end{equation}
 The radius $R_S^2$ verifies for every $i$ in $\{1, \dots, n\}$ 
\begin{equation}
\vvnorm{c_i - c_S}^2 + R_S^2 = R_i^2.
\end{equation}
\end{lemma}

\begin{proof}
Let $c_S$ be the solution of the Eq. \ref{eq:centre_sphere}. Denote by $A$ the affine space defined by equations (on $u$) $\scal{u}{(c_1 - c_k)} = K_k$ for every $2 \leq k \leq j$, to which belong any point of $S$ by subtracting equations $\vvnorm{x - c_i}^2 = R_i^2$ to $\vvnorm{x - c_1}^2 = R_1^2$. Then $c_S$ is the point at the intersection of $A$ and the affine hull of every $c_i$. For any $x$ in $A$ by the Pythagorean theorem one has $\vvnorm{x - c_i}^2 = \vvnorm{x - c_S + c_S - c_i}^2 =\vvnorm{x - c_S}^2 + \vvnorm{c_S- c_i}^2$. One can check that $R_i^2 - \vvnorm{c_i - c_S}^2$ is a positive number independent of $i$, whose square root we define to be $R_S$. Thus $x$ belongs to $S$ if and only if $x \in A$ and $\vvnorm{x - c_S}^2 = R_S^2$.
\end{proof}

\algname{\algsphereinter}
\begin{algorithm}[H]
\caption{
 Compute the center and radius of the $d- \#I$ dimensional sphere given by the intersection of transverse spheres $S_I$.}
\label{alg:SphereIntersections}
\begin{algorithmic}
\Require{A set of indices $I$ with corresponding centers $c_i$ and radii $R_i$}
\State $\lambda \gets$ solution of the system of Eqs. \ref{eq:centre_sphere}
\State $c_S \gets \sum_{i \in I} \lambda_i c_i$
\State $R_S \gets \sqrt{R_1^2 - \vvnorm{c_1 - c_S}^2}$
\State \Return{$c_S, R_S$}
\end{algorithmic}
\end{algorithm}

The second one requires a projection onto an affine hull.

\begin{lemma}[Computations in the \algminsphereinter procedure.]
Let $S = \cap_{1 \leq i \leq j} S_i$ be an intersection of $j$ spheres in $\Rd$ intersecting transversely with centers in a generic position. Let $g$ be a map of the form $x \mapsto \alpha \left ( \vvnorm{x -c}^2 - R^2 \right )$ with $c \notin S$. Then $g_{|S}$ is a smooth map whose differential vanishes on $S$ exactly at point $c_S \pm R_S \frac{b}{\vvnorm{b}}$ where $b := c - \pi(c)$, $\pi$ being the projection map onto the closest point in the affine subspace generated by the centers $c_1, \dots, c_j$. These points are the argmin and the argmax of $g_{|S}$.
\end{lemma}

\begin{proof}
Without loss of generality, take $\alpha = 1/2$ and $R=0$.
For any $x \in S$, $\nabla g_{|S}(x)$ is the orthogonal projection of $\nabla g(x) = x - c$ onto the tangent space $T_x S$. This space consists in a $d-j$ dimensional vector subspace of $\Rd$ defined by the equations (on $v$)
\begin{equation}
\scal{v}{(c_1 - c_i)} = 0 \text{ for all }2 \leq i \leq j \qquad  \scal{v}{(x - c_S)} = 0.
\end{equation}
By the change of variable $x = c_S + R_Su$ and by the previous proposition, $S$ can be parameterized by $u$ ranging in the unit sphere orthogonal to all the $c_1 - c_j$. All in all $T_x S$ is the space orthogonal to $(c_1 - c_i)_{2 \leq i \leq j}, u$. The orthogonal projection onto such a space is zero if and only if it belongs to the space spanned by these vectors. Thus the differential vanishes in $x$ if and only if 
\begin{align*}
x - c = c_S + R_S u - c \in \Vect(c_1 - c_i, u).
\end{align*}
which after some manipulation and seeing that $c_S$ is a convex combination of every $c_i$, is equivalent to 
\begin{equation}
\label{eq:appartenance}
c - c_1 \in \Vect(c_1 - c_i, u).
\end{equation}
They are exactly two unit vectors orthogonal to every $c_1 - c_i$ such that the previous equation is verified. Indeed, write $c - c_1 = a + b$ where $a \in \Vect(c_1 - c_i)$ and $b$ is orthogonal to them, which is equivalent to $b$ being $c - \pi(c)$ as defined in the proposition. The previous equation \ref{eq:appartenance} is verified if and only if $u = \pm \frac{b}{\vvnorm{b}}$. 
The argmin and the argmax being two distinct critical points of $g_{|S}$, they are necessarily these two points.
\end{proof}

\algname{\algminsphereinter}
\begin{algorithm}
\caption{Compute the minimum $y$ of a function of the type $\vvnorm{x - c}^2$ on a transverse intersection of spheres $S = \cap_{i \in I} S_i$.}
\label{alg:MinSphereIntersections}

\begin{algorithmic}
\Require{A set of indices $I$ with corresponding centers $c_i$ and radii $R_i$, a center $c$ such that $\vvnorm{x-c}^2$ is to be minimized over the intersection $S$ of the spheres indexed in $I$}
\Ensure{$y = \argmin[x\in S] \vvnorm{x - c}^2$}
\State $(c_S, R_S) \gets$ $\algsphereinter{I}$
\State $\pi(c) \gets$ the projection of $c$ into the affine hull of $(c_i)_{i \in I}$.
\State $u \gets \frac{c - \pi(c)}{\vvnorm{c - \pi(c)}}$
\State $(y_1, y_2) \gets (c_S + R_Su,c_S - R_Su)$
\If{$\vvnorm{y_1  -c}^2 \geq \vvnorm{y_2 - c}^2$}
	\Return{$y_2$}
\Else \quad \Return{$y_1$}
\EndIf
\end{algorithmic}
\end{algorithm}

The following lemma show how to compute the \algsphere procedure, that is, to compute when a geodesic $[x,y]_S$ among a sphere $S$ changes cell starting from $x$.
 
\begin{lemma}[Computations in the \algsphere procedure.]
Let $x,y$ be two non-antipodal points on a sphere of radius $R$ centered at $C$.
Then the change of signs of $f_i$ among the geodesic $[x,y]$ on the sphere are as follows.
\begin{itemize}
\item If $f_i(x)$ and $f_i(y)$ have different signs, and there is exactly one point in $[x,y]$ where $f_i$ vanishes.
\item If $f_i(x)$ and $f_i(y)$ have same signs, there is either zero or two points in $[x,y]$ where $f_i$ vanishes.
\end{itemize}
Moreover, such intersection points are computable thanks to the following process. Put $\nu : x \mapsto \frac{x}{\vvnorm{x}}$
and let $u = \nu(x - c)$ and $v = \nu(y - c - (\scal{(y - c)}{u}) u)$, forming an orthonormal basis of $\mathrm{Vect}(x-c, y-c)$. Then $S_i$ intersects with $S \cap \mathrm{Aff}(c,x,y)$ at two points if and only if 
\begin{equation}
\label{eq:check}
\vvnorm{w}^2R^2 \geq D_i^2,
\end{equation} 
where $w^x_i = \scal{(c_i - c)}{u}$ and $w^y_i = \scal{(c_i - c)}{v}$, $w_i = w^x_i u + w^y_i v$ (and thus $\vvnorm{w_i}^2 = (w^x_i)^2 + (w^y_i)^2$), $D_i = \frac{1}{2} \left ( R_i^2 - \vvnorm{c_i - c}^2 - R^2 \right )$.
If Eq. \ref{eq:check} is verified, then these two points are of the form $C + su + tv$ with 
\begin{equation}
(s,t) = \frac{D_i}{\vvnorm{w}^2}(w^x_i, w^y_i) \pm \sqrt{R^2 - \frac{D_i^2}{\vvnorm{w_i}^2}}(-w^y_i, w^y_i).
\end{equation}
Furthermore, $c + su + tv$ lies on the geodesic $[x,y]$ on $S$ if and only if $\scal{(y-c)}{u} \leq s$ and $t \geq 0$.
\end{lemma}

\begin{proof}
Since by the definition of $v$ one has $\scal{(y-c)}{v} > 0$  we have $y - c = \scal{(y-c)}{u} + \sqrt{R^2 - \scal{(y-c)}{u}^2}v$, and any point $c + su + tv$ lies on the spherical geodesic $[x,y]$ if and only if $s^2 + t^2 = R^2$ and $\scal{(y-c)}{u} \leq s, t \geq 0$. 
Now the orthogonal projection of $c_i$ onto the plane $P = \mathrm{Aff}(c,x,y)$ is obtained by $\Pi_{P}(c_i) = c + (\scal{(c_i -c)}{u})u + (\scal{(c_i - c)}{v})v = c + w_i$.
The intersection $S^P_i = S_i \cap P$ is either empty or a circle of radius $R^P_i = \sqrt{R_i^2 -  \vvnorm{c_i - c}^2 + \vvnorm{w_i}^2}$ within $P$, centered at $c + w_i$, depending on the sign of $R_i^2 -  \vvnorm{c_i - c}^2 + \vvnorm{w_i}^2$. $S^P_i$ and $S$ have non-empty intersection if and only if $\vvnorm{w_i} \leq R + R^P_i$.

Decomposed as $c + su + tv$,  the intersecting points are exactly the solutions of the quadratic system $s^2 + t^2  =  R^2,\vvnorm{su + tv - w_i}^2 = (R^P_i)^2$ which is equivalent to:
\[
\left \{ 
\begin{array}{rcl}
s^2 + t^2 & = &  R^2 \\
sw_i^x + tw_i^y & = & D_i
\end{array} \right.
\]
where $D_i  = \frac{1}{2}\left (R^2 + \vvnorm{c - c_i}^2 - R_i^2 \right)$. It is simpler to see the solution as $c + \alpha w_i + \beta w^{\perp}_i$, with $w_i^{\perp} = - w^y_i u + w^x_i v$. Then from the second equation of the last system we have $\alpha = \frac{D_i}{\vvnorm{w_i}^2}$ and $\beta = \pm \frac{1}{\vvnorm{w_i}} \sqrt{R^2 - \frac{D_i^2}{\vvnorm{w_i}^2}}$.
\end{proof}

\begin{lemma}[\algsphere decreases $F$]
Let $x$ be a point of $\Rd$ which does not minimizes $F$, with $\#I^0_*(x) \geq 2$, $g = f_{I^+_*(x)}$. As per the previous proposition let $y$ be the point of $S = S_{I^+_*(x)}$ where $g_{|S}$ reaches its minimum. Let $[x,y]_{|S}$ be the geodesic between $x$ and $y$ on $S$ (or any geodesic if $x$ and $y$ are antipodal). Let $\lambda$ be the minimum of $t$ such that the projection $x_t$ of $tx + (1-t)y$ on $S$ (or equivalently, onto $[x,y]_S$) has one of $I^+(x_t)$, $I^-(x_t)$ different from respectively $I^+_*(x)$ and $I^-_*(x)$. Then $\lambda$ is well-defined and we have 
\begin{equation}
F(x_{\lambda}) < F(x).
\end{equation}
\end{lemma} 

\begin{proof}
By Lemma \ref{lem:neighboring_cells}, we know that $x$ belongs to the closure of $\calC^*(x)$. Moreover for $t$ small enough, $x_t$ belongs to this cell as by definition $x_t$ stays on $S$. Now on the trajectory $t \mapsto x_t$, $F$ coincides with $f_{I^+_*}(x)$ which is a strictly convex function. Its restriction to $S$ has a unique minimum attained at $y$; thus $F$ it decreases along the geodesic $[x,y]_S$. 
\end{proof}

\algname{\algsphere}
\begin{algorithm}[h!]
\caption{Compute the first point along the geodesic between $x,y$ on a Sphere $S_{I^0}$ of center $C$, radius $R$ leaving the cell and the cell to which this point belong.}
\label{alg:SphereDescent}
\begin{algorithmic}
\Require{A set of centers $c_i$ and radii $R_i$, an initial point $x$, destination $y$, sets of indices $I^+, I^0, I^-$.}
\Ensure{$z = C + su + \sqrt{R^2 - s^2} v$ is the first point along the geodesic between $x,y$ on a sphere leaving the cell.} 
\State $(s,t) \gets (-R,0)$
\State $u \gets \frac{x - C}{\vvnorm{x- C}}$
\State $v \gets y - C - (y - C) \cdot u$
\State $v \gets \frac{v}{\vvnorm{v}}$
\State ChangingIndices $\gets\emptyset$ \Comment{{\togray Set of indices in $I^+ \cup I^-$ to which the final $z$ belongs - usually a singleton.}}
\For{$i \in I^+ \cup I^-$}
 \State $D_i \gets \frac{1}{2}\left (R_i^2 - \vvnorm{c_i - C}^2 - R^2 \right )$
 \State $w^x_i \gets (c_i - C) \cdot u, w^y_i \gets (c_i - C) \cdot v$
 \State $w_i \gets (w^x_i)^2 + (w^y_i)^2$
 \State $\Delta_i \gets  R^2 - \frac{D_i^2}{w_i}$
  \If{$\Delta_i > 0$} \Comment{{ \togray If the circle on which the geodesic lies intersects with $S_i$}}
    \State $s_+ \gets \frac{D_i}{w_i} w^x_{i} + \sqrt{\frac{\Delta_i}{w_i}}w^y_{i}, s_- \gets  \frac{D_i}{w_i} w^x_{i} - \sqrt{\frac{\Delta_i}{w_i}}w^y_{i}$
    \State $t_+ \gets \frac{D_i}{w_i} w^y_{i} - \sqrt{\frac{\Delta_i}{w_i}}w^x_{i}, t_- \gets  \frac{D_i}{w_i} w^x_{i} + \sqrt{\frac{\Delta_i}{w_i}}w^x_{i}$
    \State $L \gets$ set of pairs among $(s_+, t_+),(s_-, t_-)$ verifying $s_{\cdot} \geq s, t_{\cdot} \geq 0$.
    \State $(s_{temp}, t_{temp}) \gets$ the pair with maximal first element in $L$. 
    \If{ $s_{temp} = s$} \Comment{{ \togray Degenerate case where $z = c + su + tv$ (at that point in the algorithm) already lies on a sphere $S_i$ with $i \notin I^0$.}}     
    \State Add $i$ to ChangingIndices
    \Else 
    \State ChangingIndices = $[i]$
	\State $(s,t) \gets (s_{temp}, t_{temp})$
    \EndIf 
   \EndIf 
 \EndFor 
 \State Add elements of ChangingIndices to $I^0$, Remove elements from ChangingIndices off of $I^+$, $I^-$.
\State \Return{$z = C + su + tv$, $I^0$, $I^+$, $I^-$}
\end{algorithmic}
\end{algorithm}

\subsection{Exact algorithm : pseudocode and convergence result}

\algname{\algexact}
\vspace{-1em}
\begin{algorithm}[H]

\begin{algorithmic}
\Require{A set of centers $c_i$ and radii $R_i$, an initial point $x_0$}
\Ensure{$x$ is the point where the minimum of $F$ is attained} 
\State $x \gets x_0$, MinimumAttained $\gets$ False.
\State $I^+(x), I^0(x), I^-(x) \gets \sgns(x)$
\While{MinimumAttained is False}
\If{$\#I^0(x)$}\label{alg:kzero-done}
	
\If{\textcolor{teal}{$c_{I^+(x)} \in \calC(x)$}} \Comment{\textcolor{gray}{Check if the minimum lies in $\calC(x)$.}}
\State 	\algtele procedure : MinimumAttained $\gets$ True, $x \gets c_{I^+(x)}$.
\Else
\State	\LComment{ \color{red} Compute the lowest $t$ such that $x -t\nabla f_{I^+(x)}$ is in another cell using the \hyperref[alg:LineDescent]{\algline} algorithm and update the signs. \label{alg:dim_pleine_départ}}
\State	$x, I^+, I^0, I^- \gets$ \algline($x,- \nabla f_{I^+(x)}$).
\EndIf 

\Else \Comment{\textcolor{gray}{$x$ lies on at least one $S_i$}} \label{alg:codim_pos}
\State \LComment{\color{violet} Compute $\nabla_* F(x)$ and the sets of indices $I^+_*(x), I^0_*(x), I^-_*(x)$ via \algqp}
\State $I^+_*(x), I^0_*(x), I^-_*(x) \gets \algqp(x, I^+(x), I^0(x), I^-(x)).$ 
\If{\textcolor{violet}{$0 \in \clarke  F(x)$}} \Comment{\textcolor{gray}{Check if the minimum is attained at $x$.}}
\State	MinimumAttained $\gets$ True \label{alg:fin}
\ElsIf{$\#I^0_*(x) =0$ } \Comment{\textcolor{gray}{Traj. out of $x$ is in a full dimensional cell.}}

\If{\textcolor{teal}{$c_{I^+_*(x)} \in \calC^*(x)$}} \Comment{\textcolor{gray}{Check if the minimum lies in $\calC^*(x)$.}}
\State \algtele procedure : $x \gets c_{I^+_*(x)}$, MinimumAttained $\gets$ True
\Else 
\State \LComment{ \color{red} Compute the lowest $t$ such that $x -t\nabla f_{I^+_*(x)}$ is in another cell using the  \hyperref[alg:LineDescent]{\algline} algorithm and update the signs.}
\State $x, I^+, I^0, I^- \gets$ \algline($x,- \nabla f_{I^+_*(x)}$)  \label{alg:dim_pleine_traj}
\EndIf

\Else \Comment{\textcolor{gray}{Trajectory out of $x$ is \emph{not} in a full dimensional cell.}} \label{alg:codim_pos2}
\State \LComment{\color{olive} Compute $c_S, R_S$ respectively the center/radius of $S_{I^0_*(x)}$ via \algsphereinter.} 
\State $c_S, R_S \gets$ \algsphereinter$(I^0_*(x))$.
\State \LComment{\color{olive} Compute the argmin of $f_{|I^+_*(x)}$ on $S_{I^0_*(x)}$ using the \hyperref[alg:MinSphereIntersections]{\algminsphereinter} algorithm.}\label{alg:minimum}
\State $y \gets $ \algminsphereinter$(I^+_*(x), c_S, R_S)$
\If{\textcolor{teal}{$y \in \calC^*(x)$}}
\State $x \gets y$
\Else 
\State \LComment{\color{red} Compute the first intersection of the geodesic from $x$ to $y$ on $S$ with another cell using the \hyperref[alg:SphereDescent]{\algsphere} algorithm. Update the signs. \label{alg:codim_pos_traj}}
\State	$x, I^+, I^0, I^- \gets $ \algsphere($x,y,S$)
\EndIf
\EndIf 
\EndIf 
\EndWhile
\State \Return{$x$}
\end{algorithmic}
\caption{Compute the minimum of $F = \sum_{i} \max(0, \vvnorm{x-c_i}^2 - R_i^2)$.}
\label{alg:exact}
\vspace{0.1em}
\hrule
\vspace{0.5em}
\textcolor{red}{Red} = Computations taking the minimum of roots of second degree polynomials. \\
\textcolor{violet}{Violet} = Computing the solution of a quadratic programming problem. \\
\textcolor{olive}{Olive} = Computations solving a full-rank linear system and roots of degree two polynomials.\\
\textcolor{teal}{Teal} = Exact computations of predicates.
\end{algorithm}

\begin{theorem}[Algorithm convergence]
There exists a neighborhood of the point $x^*$ where $F$ reaches its minimum, such that for any starting point in this neighborhood the algorithm converges in at most $3^c$ operations, where $c$ is the number of spheres on which $x^*$ lies.
\end{theorem}

\textit{Proof sketch.}
At point $x^*$, the minimum of $F$ is reached meaning that its Clarke gradient $\clarke F(x^*)$ contains 0. Recall that this Clarke gradient $\partial_* F(x)$ of $F$ at any point $x$ is a box of the form
\[ \{ \nabla f_{I^+(x)}(x) + \sum_{i \in I^0(x)} \alpha_i \nabla f_i(x), 0 \leq \alpha_i \leq 1 \}\]
From classical geometric combinatorics, such a box has $3^c$ facets, where $c$ is the cardinal of $I^0(x)$. For any point $x$ with $I^0(x) \subset I^0(x^*)$, $I^+(x^*) \subset I^+(x),  I^+(x^*) \subset I^+(x)$, $\partial_* F(x)$ is a box of smaller dimension than $\partial_* F(x^*)$, and it is similar to a facet of the box $\partial_* F(x^*)$. 

The underlying trajectory followed by our algorithm takes direction in the Clarke gradient of $F$, so that it actually gets closer to $x^*$. Now take a (small enough) neighborhood $U$ of $x^*$ so that:
\begin{itemize}
\item the cells intersecting with $U$ are exactly the cells whose adherence contains $x^*$;
\item Any semiflow trajectory starting from a point in $U$ stays in $U$ (typically think of $U$ as a ball centered in $x^*$, thanks to the previous remark).
\end{itemize}

By the convex nature of the functional our underlying semiflow trajectory has necessarily decreasing norm of generalized gradient - it is indeed following either the flow of a convex function on a submanifold, or branching when computing the Clarke gradient to a new generalized gradient with even smaller norm. Considering $x_n$ the sequence of elements obtained by our algorithm starting from any point in $U$, this means that the sequence $\vvnorm{\nabla_* F(x_n)}$ is strictly decreasing. 

Since the Clarke gradients $\clarke F(x_n)$ are close\footnote{In the sense that the vectors defining the boxes at points $x_n, x^*$ are at distance smaller than $2\vvnorm{x^* - x_n}$} to facets of the box $\clarke F(x^*)$, this suffices to obtain convergence. Ordering the facets by the distance of 0 to said facets yields shows that the sequence of facets of $\partial_* F(x)$ associated to $\partial_* F(x_n)$ is strictly decreasing, and must thus converge to the full box.

\clearpage
\section{Experiments}
\label{sec:exp}

\subsection{Datasets clustering}

\begin{table}[!ht] 
\begin{center}
\begin{tabular}{|l r|}
\hline
Name & Shape $(n,d)$\\
\hline
yeast & $(1484, 9)$\\
faults & $(1941, 28)$\\
mfeat & $(2000, 77)$\\
cloud & $(2048, 11)$\\
segmentation & $(2310, 20)$\\
spam & $(4601, 58)$\\
optdigits & $(5620, 65)$\\
landsat & $(6435, 37)$\\
pendigits& $(10992, 17)$\\
letter& $(20000, 17)$ \\
\hline
\end{tabular}
\end{center}
\caption{{\bf The dataset $\datasetCLU$, consisting of 10 individual datasets.}
The shape of a dataset: $n$: number of observations/points; $d$: dimension.
Datasets selected in clustering experiments, see \cite{celebi2013comparative}.
} 
\label{tab:datasetCLU} 
\end{table} 

\begin{table}[!ht] 
\begin{center}
\begin{tabular}{| rr | rr | rr| rr |}
\hline
id & $n$ & id & $n$ & id & $n$ & id & $n$\\
\hline
cluster\_0 &     20 & cluster\_1 &       19 & cluster\_2 &       11 & cluster\_3 &       49\\
cluster\_4 &       45 & cluster\_5 &       77 & cluster\_6 &       69 & cluster\_7 &      101\\
cluster\_8 &       90 & cluster\_9 &      161 & cluster\_10 &      149 & cluster\_11 &      176 \\
cluster\_12 &      100 & cluster\_13 &       92 & cluster\_14 &      165 & cluster\_15 &      119 \\
\hline
\end{tabular}
\end{center}
\caption{{\bf The \datasetHMM dataset: 16 clusters in dimension $d=400$ -- from \cite{vicedomini2022profileview}.}
 We use the clusters as individual datasets.}
\label{tab:datasetHMM} 
\end{table} 

\clearpage

\subsubsection{Dataset \datasetCLU: comparison \algoexact vs \algobfgs}

\begin{table}[htb]
\begin{center}
\begin{tabular}{|l|rrr|}
\hline
 &  min  &  median  &  max\\
\hline
$\eta=0.10$ &  2.694e-04  &  1.001e-03  &  2.875e-03\\
$\eta=0.20$ &  3.795e-04  &  6.018e-03  &  8.253e-02\\
$\eta=0.30$ &  2.101e-04  &  4.999e-02  &  2.583e-01\\
$\eta=0.40$ &  1.554e-04  &  1.112e-01  &  5.133e-01\\
$\eta=0.50$ &  5.921e-04  &  1.635e-01  &  1.160e+00\\
$\eta=0.60$ &  3.834e-03  &  3.567e-01  &  1.608e+00\\
$\eta=0.70$ &  2.143e-02  &  7.780e-01  &  8.065e+00\\
$\eta=0.80$ &  7.910e-02  &  1.076e+00  &  1.043e+01\\
$\eta=0.90$ &  9.749e-02  &  1.252e+00  &  1.236e+01\\
\hline
\end{tabular}
\end{center}
\caption{{\bf DatasetCLU: $\timeexact / \timebfgs$.}}
\label{tab:DatasetCLU-times-exact-BFGS}
\end{table}

\begin{table}[htb]
\begin{center}
\begin{tabular}{|l|rrr|}
\hline
 &  min  &  median  &  max\\
\hline
$\eta=0.10$ &  1.000e+00  &  1.000e+00  &  1.000e+00\\
$\eta=0.20$ &  1.000e+00  &  1.000e+00  &  1.000e+00\\
$\eta=0.30$ &  1.000e+00  &  1.000e+00  &  1.000e+00\\
$\eta=0.40$ &  1.000e+00  &  1.000e+00  &  1.000e+00\\
$\eta=0.50$ &  1.000e+00  &  1.000e+00  &  1.000e+00\\
$\eta=0.60$ &  1.000e+00  &  1.000e+00  &  1.000e+00\\
$\eta=0.70$ &  1.000e+00  &  1.000e+00  &  1.000e+00\\
$\eta=0.80$ &  1.000e+00  &  1.000e+00  &  1.000e+00\\
$\eta=0.90$ &  1.000e+00  &  1.000e+00  &  1.000e+00\\
\hline
\end{tabular}
\end{center}
\caption{{\bf DatasetCLU: $\Feta{\optexact} / \Feta{\optbfgs}$.}}
\label{tab:DatasetCLU-values-exact-BFGS}
\end{table}

\FloatBarrier

\clearpage
\subsubsection{Dataset \datasetCLU: comparison \algoexact vs \algolbfgs}

\begin{table}[htb]
\begin{center}
\begin{tabular}{|l|rrr|}
\hline
 &  min  &  median  &  max\\
\hline
$\eta=0.10$ &  1.660e-03  &  8.070e-03  &  5.783e-02\\
$\eta=0.20$ &  1.350e-03  &  1.403e-01  &  2.605e+00\\
$\eta=0.30$ &  2.060e-03  &  7.913e-01  &  3.208e+00\\
$\eta=0.40$ &  7.021e-04  &  1.304e+00  &  7.195e+00\\
$\eta=0.50$ &  1.394e-02  &  1.521e+00  &  9.129e+00\\
$\eta=0.60$ &  2.027e-01  &  3.627e+00  &  3.103e+01\\
$\eta=0.70$ &  5.270e-01  &  2.777e+00  &  5.143e+01\\
$\eta=0.80$ &  6.523e-01  &  6.386e+00  &  5.227e+01\\
$\eta=0.90$ &  5.829e-01  &  5.068e+00  &  6.467e+01\\
\hline
\end{tabular}
\end{center}
\caption{{\bf DatasetCLU: $\timeexact / \timelbfgs$.}}
\label{tab:DatasetCLU-times-exact-LBFGS}
\end{table}

\begin{table}[htb]
\begin{center}
\begin{tabular}{|l|rrr|}
\hline
 &  min  &  median  &  max\\
\hline
$\eta=0.10$ &  1.000e+00  &  1.000e+00  &  1.000e+00\\
$\eta=0.20$ &  1.000e+00  &  1.000e+00  &  1.000e+00\\
$\eta=0.30$ &  1.000e+00  &  1.000e+00  &  1.000e+00\\
$\eta=0.40$ &  1.000e+00  &  1.000e+00  &  1.000e+00\\
$\eta=0.50$ &  1.000e+00  &  1.000e+00  &  1.000e+00\\
$\eta=0.60$ &  1.000e+00  &  1.000e+00  &  1.000e+00\\
$\eta=0.70$ &  1.000e+00  &  1.000e+00  &  1.000e+00\\
$\eta=0.80$ &  1.000e+00  &  1.000e+00  &  1.000e+00\\
$\eta=0.90$ &  1.000e+00  &  1.000e+00  &  1.000e+00\\
\hline
\end{tabular}
\end{center}
\caption{{\bf DatasetCLU: $\Feta{\optexact} / \Feta{\optlbfgs}$.}}
\label{tab:DatasetCLU-values-exact-LBFGS}
\end{table}

\FloatBarrier

\subsubsection{Illustration: \datasetCLU, dataset yeast\_std}

\begin{figure}[htb]
\begin{center}
\begin{tabular}{ccc}
\includegraphics[width=0.306\textwidth]{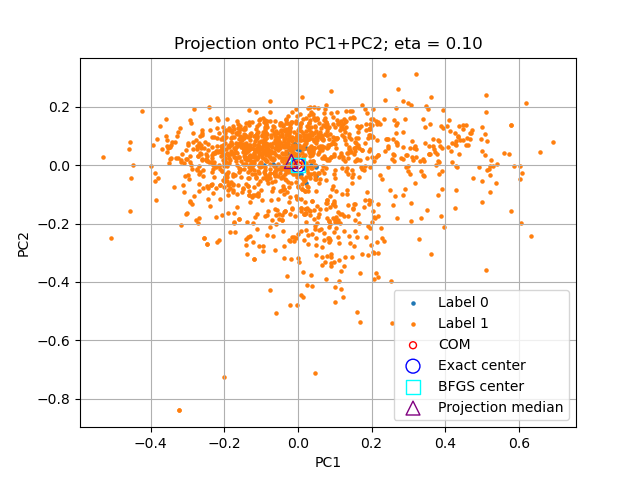} & \includegraphics[width=0.306\textwidth]{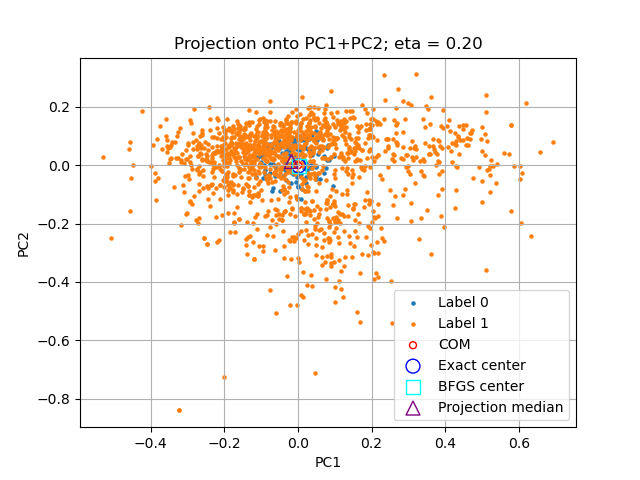} & \includegraphics[width=0.306\textwidth]{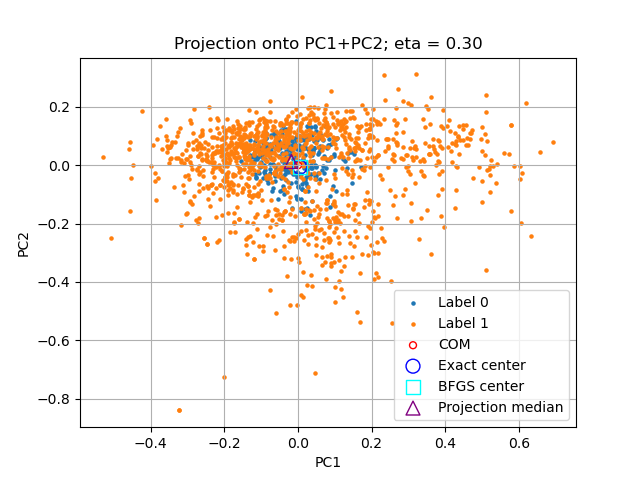}\\
{\scriptsize Projection plot yeast\_std-, $\eta=0.1$} &{\scriptsize Projection plot yeast\_std-, $\eta=0.2$} &Projection plot yeast\_std-, $\eta=0.3$\\
\includegraphics[width=0.306\textwidth]{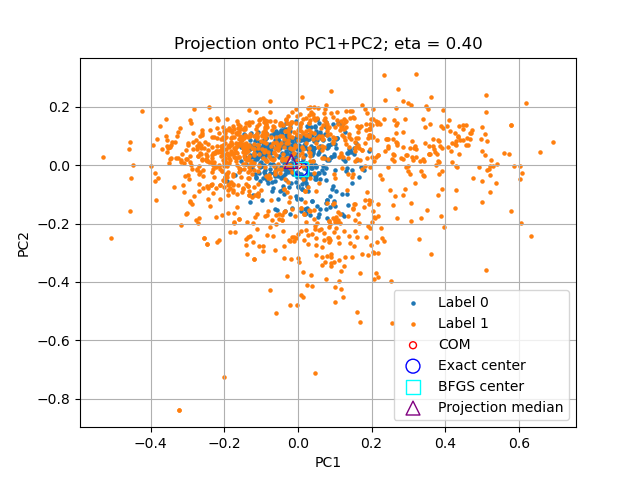} & \includegraphics[width=0.306\textwidth]{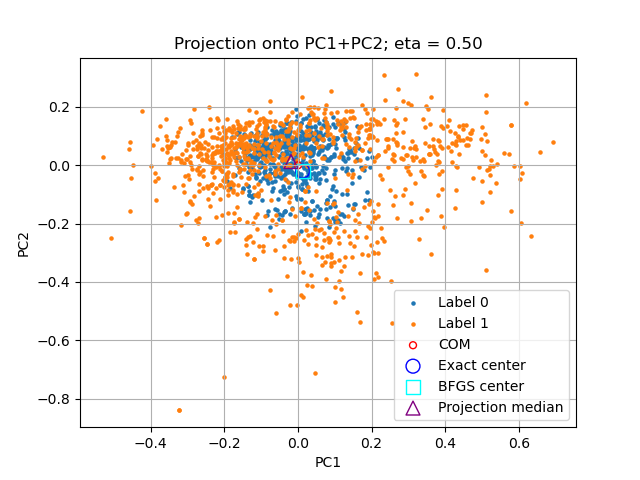} & \includegraphics[width=0.306\textwidth]{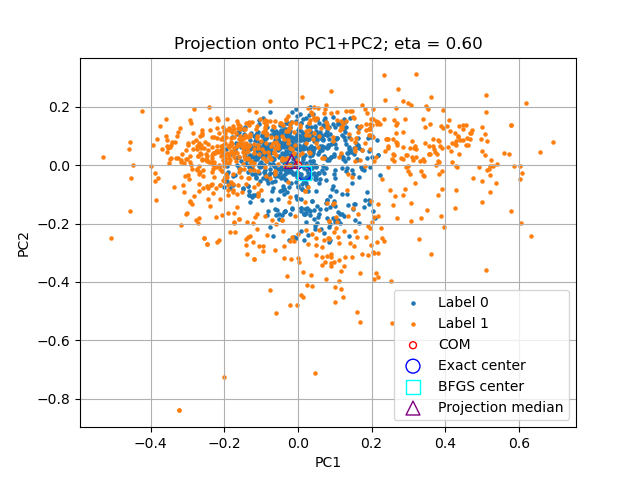}\\
{\scriptsize Projection plot yeast\_std-, $\eta=0.4$} &{\scriptsize Projection plot yeast\_std-, $\eta=0.5$} &Projection plot yeast\_std-, $\eta=0.6$\\
\includegraphics[width=0.306\textwidth]{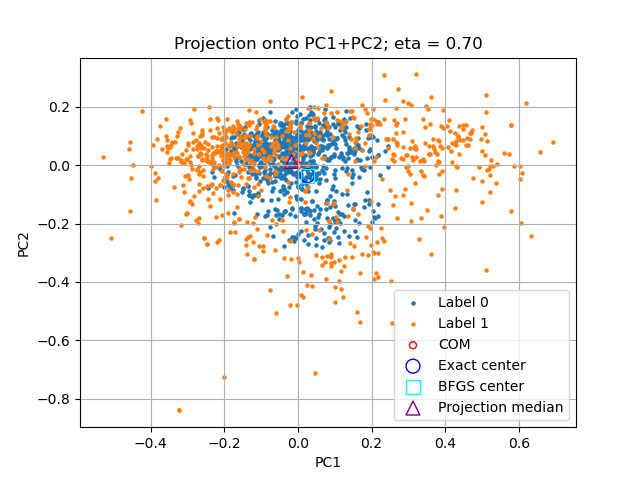} & \includegraphics[width=0.306\textwidth]{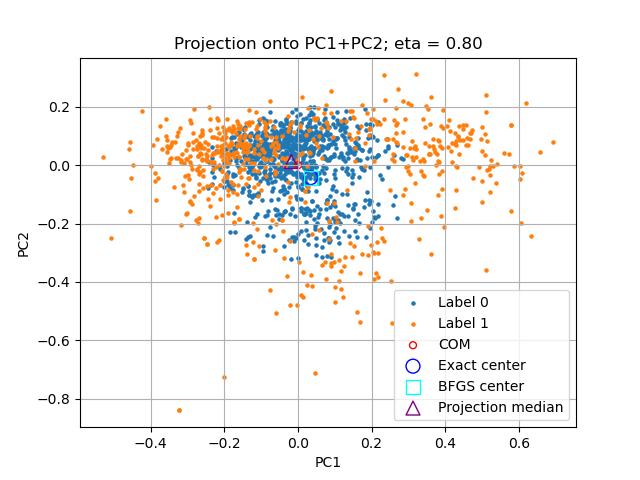} & \includegraphics[width=0.306\textwidth]{yeast_std--sescmu1dot00-sesceta0dot90--randinit0-projection-plot.png}\\
{\scriptsize Projection plot yeast\_std-, $\eta=0.7$} &{\scriptsize Projection plot yeast\_std-, $\eta=0.8$} &{\scriptsize Projection plot yeast\_std-, $\eta=0.9$}
\end{tabular}
\end{center}
\caption{{\bf \datasetCLU, dataset yeast\_std -- shape (1484,  9)}}
\label{fig:pp-yeast_std}
\end{figure}
\FloatBarrier

\subsubsection{Illustration: \datasetCLU, dataset yeast\_std}

\begin{figure}[htb]
\begin{center}
\begin{tabular}{ccc}
\includegraphics[width=0.306\textwidth]{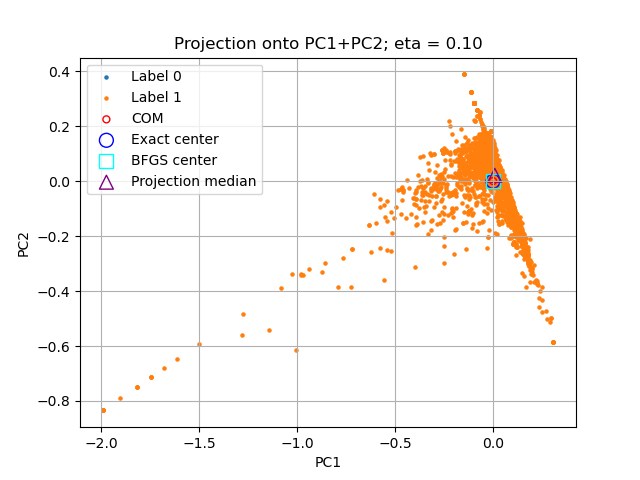} & \includegraphics[width=0.306\textwidth]{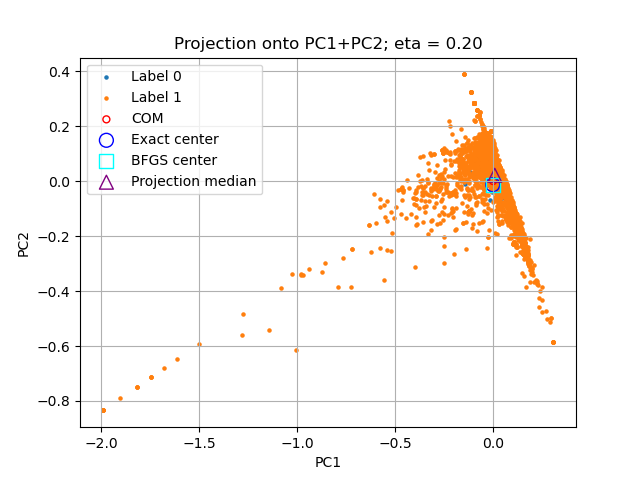} & \includegraphics[width=0.306\textwidth]{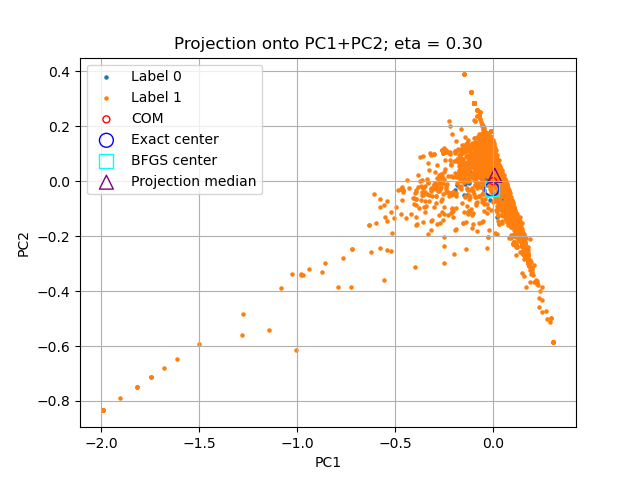}\\
{\scriptsize Projection plot spam\_std-, $\eta=0.1$} &{\scriptsize Projection plot spam\_std-, $\eta=0.2$} &Projection plot spam\_std-, $\eta=0.3$\\
\includegraphics[width=0.306\textwidth]{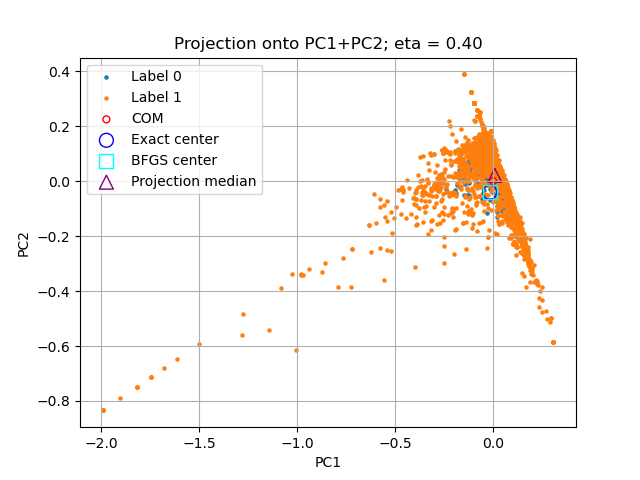} & \includegraphics[width=0.306\textwidth]{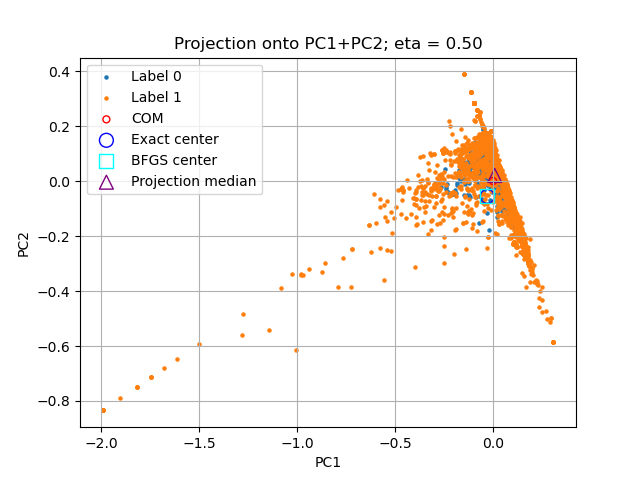} & \includegraphics[width=0.306\textwidth]{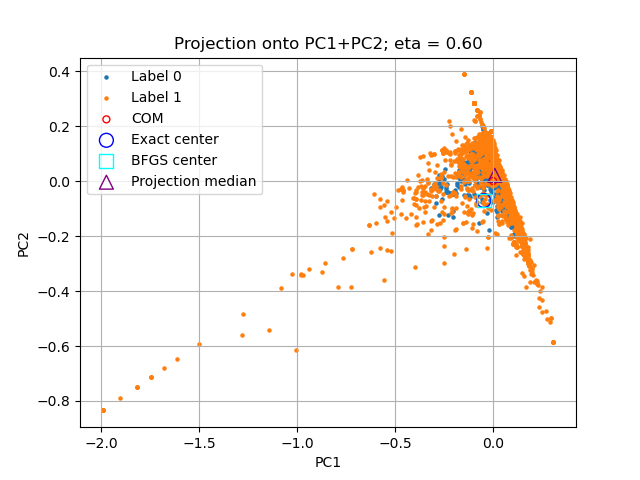}\\
{\scriptsize Projection plot spam\_std-, $\eta=0.4$} &{\scriptsize Projection plot spam\_std-, $\eta=0.5$} &Projection plot spam\_std-, $\eta=0.6$\\
\includegraphics[width=0.306\textwidth]{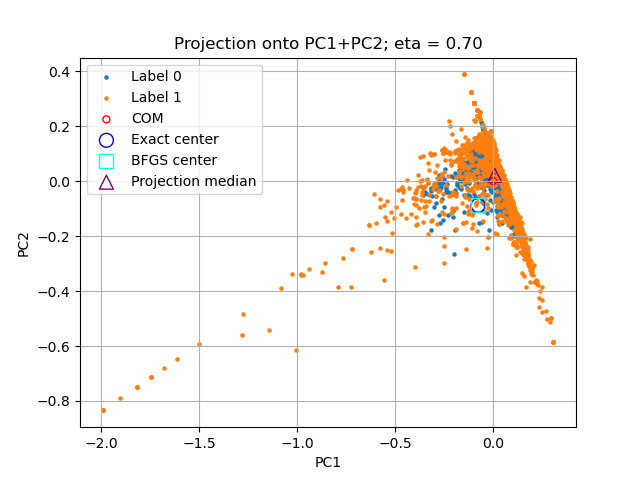} & \includegraphics[width=0.306\textwidth]{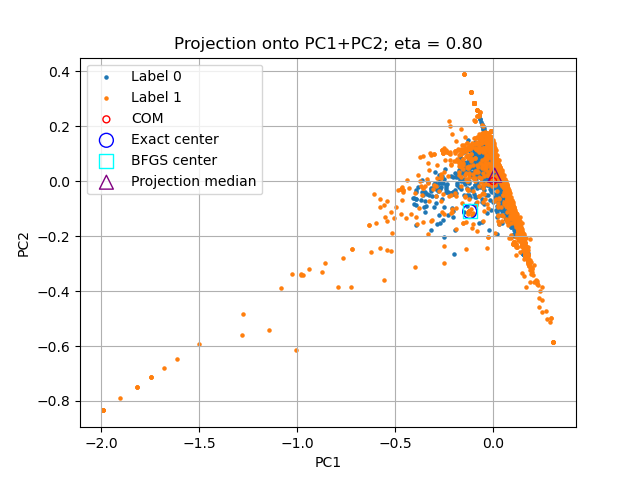} & \includegraphics[width=0.306\textwidth]{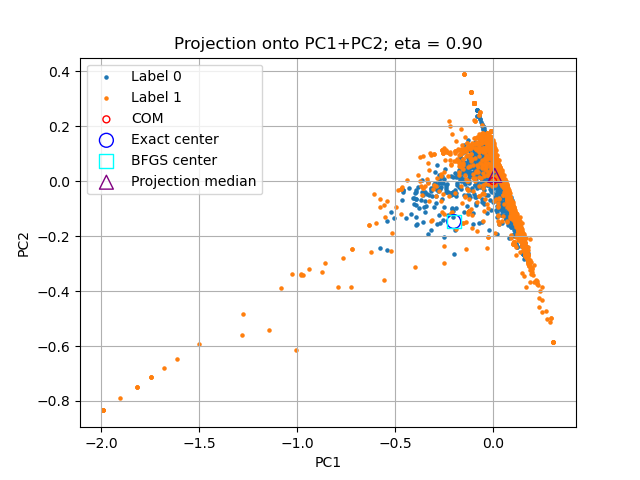}\\
{\scriptsize Projection plot spam\_std-, $\eta=0.7$} &{\scriptsize Projection plot spam\_std-, $\eta=0.8$} &{\scriptsize Projection plot spam\_std-, $\eta=0.9$}
\end{tabular}
\end{center}
\caption{{\bf Dataset spam\_std -- shape (4601,  58)}}
\label{fig:pp-spam_std}
\end{figure}
\FloatBarrier

\clearpage
\subsection{Dataset \datasetHMM}

\subsubsection{\datasetHMM: comparison \algoexact vs \algobfgs}

\clearpage
\begin{table}[htb]
\begin{center}
\begin{tabular}{|l|rrr|}
\hline
 &  min  &  median  &  max\\
\hline
$\eta=0.10$ &  2.355e-05  &  1.253e-04  &  2.194e-03\\
$\eta=0.20$ &  1.585e-05  &  1.434e-04  &  2.318e-03\\
$\eta=0.30$ &  1.021e-06  &  1.457e-04  &  1.723e-03\\
$\eta=0.40$ &  4.014e-07  &  1.507e-04  &  1.834e-03\\
$\eta=0.50$ &  1.213e-06  &  2.618e-04  &  3.800e-03\\
$\eta=0.60$ &  5.484e-04  &  1.088e-03  &  5.961e-03\\
$\eta=0.70$ &  1.460e-03  &  3.091e-03  &  2.425e-02\\
$\eta=0.80$ &  9.693e-04  &  4.121e-03  &  2.256e-02\\
$\eta=0.90$ &  2.363e-03  &  8.852e-03  &  4.387e-02\\
\hline
\end{tabular}
\end{center}
\caption{{\bf ProtHMM: $\timeexact / \timebfgs$.}}
\label{tab:ProtHMM-times-exact-BFGS}
\end{table}

\begin{table}[htb]
\begin{center}
\begin{tabular}{|l|rrr|}
\hline
 &  min  &  median  &  max\\
\hline
$\eta=0.10$ &  1.000e+00  &  1.000e+00  &  1.000e+00\\
$\eta=0.20$ &  1.000e+00  &  1.000e+00  &  1.000e+00\\
$\eta=0.30$ &  1.000e+00  &  1.000e+00  &  1.000e+00\\
$\eta=0.40$ &  1.000e+00  &  1.000e+00  &  1.000e+00\\
$\eta=0.50$ &  1.000e+00  &  1.000e+00  &  1.000e+00\\
$\eta=0.60$ &  9.999e-01  &  1.000e+00  &  9.999e-01\\
$\eta=0.70$ &  9.992e-01  &  1.000e+00  &  9.992e-01\\
$\eta=0.80$ &  9.975e-01  &  1.000e+00  &  9.975e-01\\
$\eta=0.90$ &  9.892e-01  &  1.000e+00  &  9.892e-01\\
\hline
\end{tabular}
\end{center}
\caption{{\bf ProtHMM: $\Feta{\optexact} / \Feta{\optbfgs}$.}}
\label{tab:ProtHMM-values-exact-BFGS}
\end{table}

\FloatBarrier

\subsubsection{\datasetHMM: comparison \algoexact vs \algolbfgs}

\clearpage

\begin{table}[htb]
\begin{center}
\begin{tabular}{|l|rrr|}
\hline
 &  min  &  median  &  max\\
\hline
$\eta=0.10$ &  5.327e-05  &  1.287e-04  &  1.508e-03\\
$\eta=0.20$ &  7.769e-05  &  1.436e-04  &  7.514e-04\\
$\eta=0.30$ &  6.607e-05  &  2.455e-04  &  1.695e-03\\
$\eta=0.40$ &  6.549e-05  &  2.572e-04  &  9.777e-03\\
$\eta=0.50$ &  9.009e-05  &  2.133e-03  &  1.386e-02\\
$\eta=0.60$ &  1.740e-03  &  7.108e-03  &  3.145e-02\\
$\eta=0.70$ &  7.868e-03  &  2.164e-02  &  6.013e-02\\
$\eta=0.80$ &  1.322e-02  &  4.556e-02  &  1.233e-01\\
$\eta=0.90$ &  4.903e-02  &  8.999e-02  &  1.744e-01\\
\hline
\end{tabular}
\end{center}
\caption{{\bf ProtHMM: $\timeexact / \timelbfgs$.}}
\label{tab:ProtHMM-times-exact-LBFGS}
\end{table}

\begin{table}[htb]
\begin{center}
\begin{tabular}{|l|rrr|}
\hline
 &  min  &  median  &  max\\
\hline
$\eta=0.10$ &  1.000e+00  &  1.000e+00  &  1.000e+00\\
$\eta=0.20$ &  1.000e+00  &  1.000e+00  &  1.000e+00\\
$\eta=0.30$ &  1.000e+00  &  1.000e+00  &  1.000e+00\\
$\eta=0.40$ &  1.000e+00  &  1.000e+00  &  1.000e+00\\
$\eta=0.50$ &  1.000e+00  &  1.000e+00  &  1.000e+00\\
$\eta=0.60$ &  1.000e+00  &  1.000e+00  &  1.000e+00\\
$\eta=0.70$ &  9.991e-01  &  1.000e+00  &  9.991e-01\\
$\eta=0.80$ &  9.962e-01  &  9.999e-01  &  9.962e-01\\
$\eta=0.90$ &  9.232e-01  &  9.987e-01  &  9.232e-01\\
\hline
\end{tabular}
\end{center}
\caption{{\bf ProtHMM: $\Feta{\optexact} / \Feta{\optlbfgs}$.}}
\label{tab:ProtHMM-values-exact-LBFGS}
\end{table}

\FloatBarrier

\clearpage
\subsubsection{Illustration: \datasetHMM, Cluster 10}

\begin{figure}[htb]
\begin{center}
\begin{tabular}{ccc}
\includegraphics[width=0.306\textwidth]{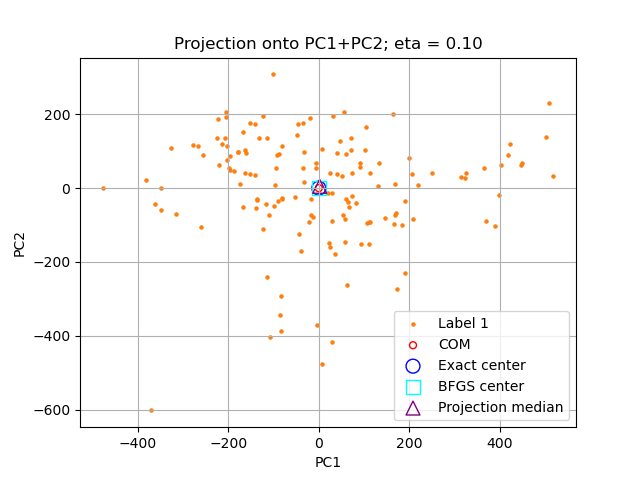} & \includegraphics[width=0.306\textwidth]{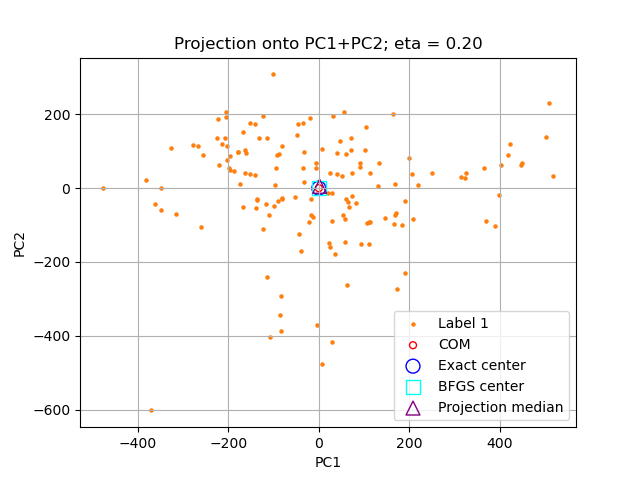} & \includegraphics[width=0.306\textwidth]{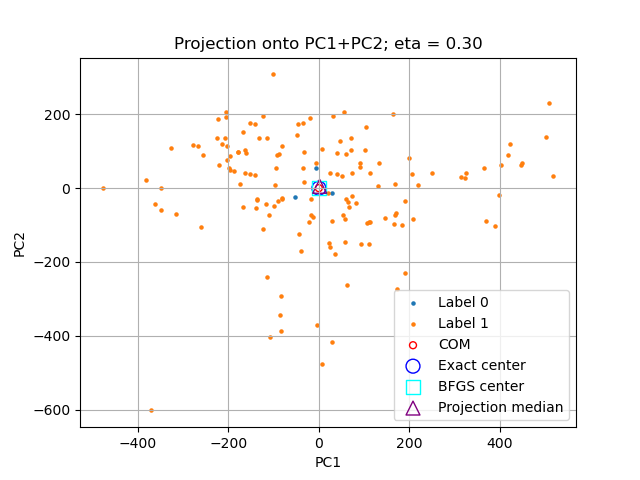}\\
{\scriptsize Projection plot cluster\_10-, $\eta=0.1$} &{\scriptsize Projection plot cluster\_10-, $\eta=0.2$} &Projection plot cluster\_10-, $\eta=0.3$\\
\includegraphics[width=0.306\textwidth]{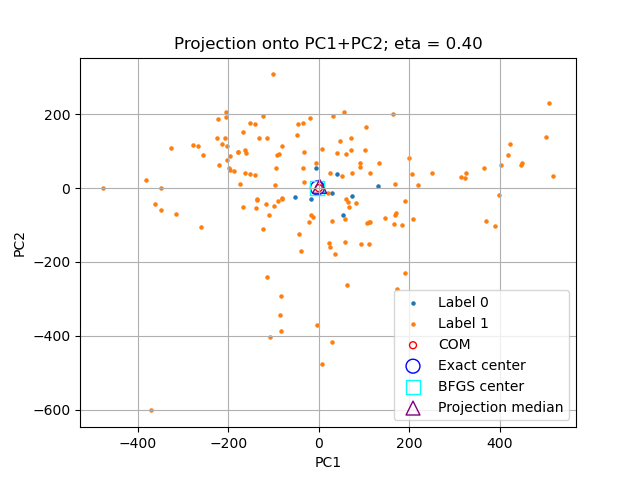} & \includegraphics[width=0.306\textwidth]{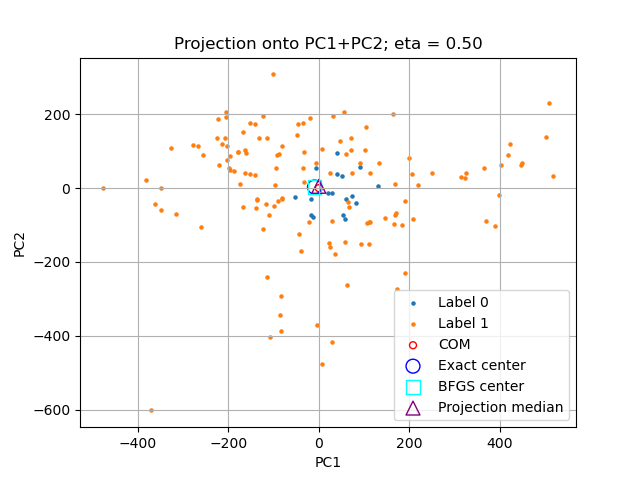} & \includegraphics[width=0.306\textwidth]{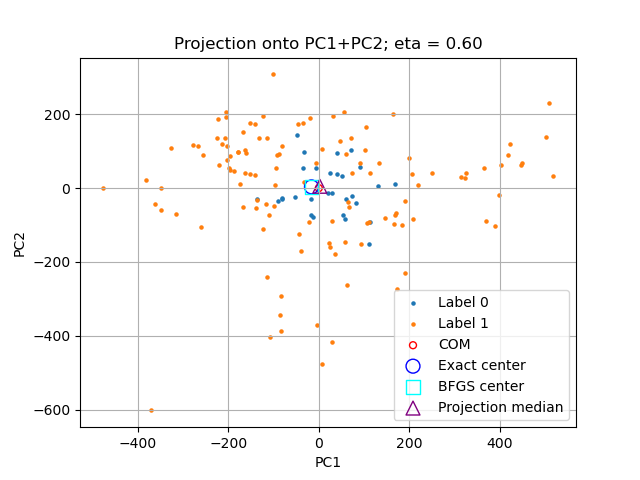}\\
{\scriptsize Projection plot cluster\_10-, $\eta=0.4$} &{\scriptsize Projection plot cluster\_10-, $\eta=0.5$} &Projection plot cluster\_10-, $\eta=0.6$\\
\includegraphics[width=0.306\textwidth]{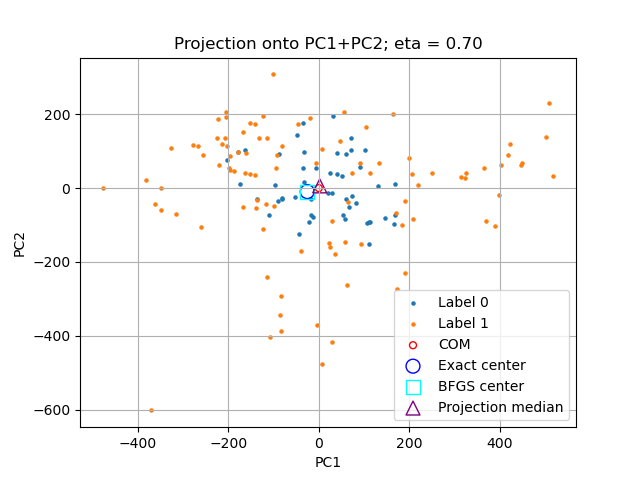} & \includegraphics[width=0.306\textwidth]{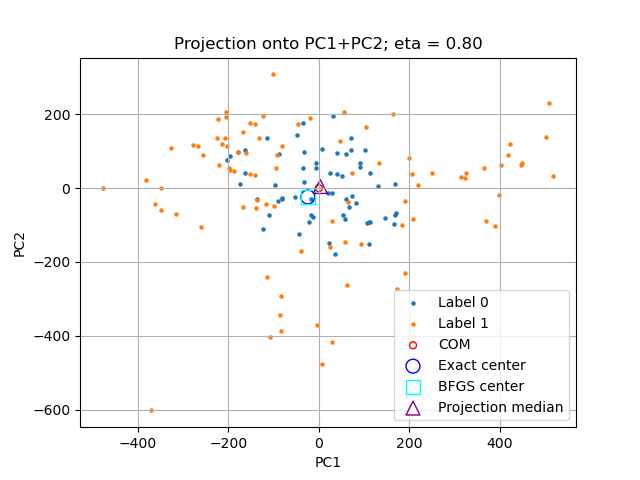} & \includegraphics[width=0.306\textwidth]{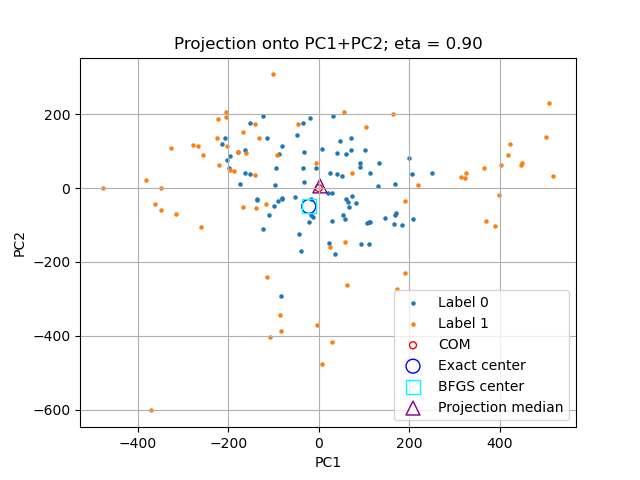}\\
{\scriptsize Projection plot cluster\_10-, $\eta=0.7$} &{\scriptsize Projection plot cluster\_10-, $\eta=0.8$} &{\scriptsize Projection plot cluster\_10-, $\eta=0.9$}
\end{tabular}
\end{center}
\caption{{\bf Dataset cluster\_10 -- shape (149,  400)}}
\label{fig:pp-cluster_10}
\end{figure}
\FloatBarrier

\clearpage

\paragraph{Illustration: \datasetHMM, Cluster 11}

\begin{figure}[htb]
\begin{center}
\begin{tabular}{ccc}
\includegraphics[width=0.306\textwidth]{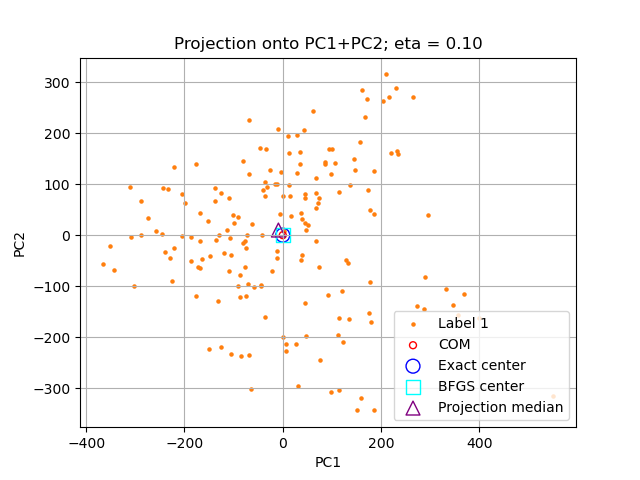} & \includegraphics[width=0.306\textwidth]{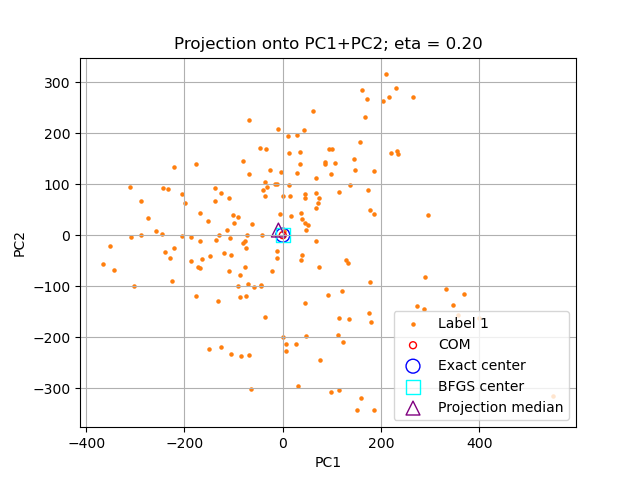} & \includegraphics[width=0.306\textwidth]{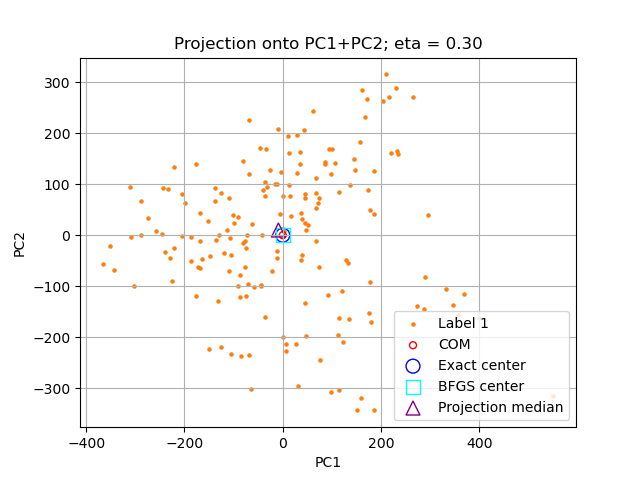}\\
{\scriptsize Projection plot cluster\_11-, $\eta=0.1$} &{\scriptsize Projection plot cluster\_11-, $\eta=0.2$} &Projection plot cluster\_11-, $\eta=0.3$\\
\includegraphics[width=0.306\textwidth]{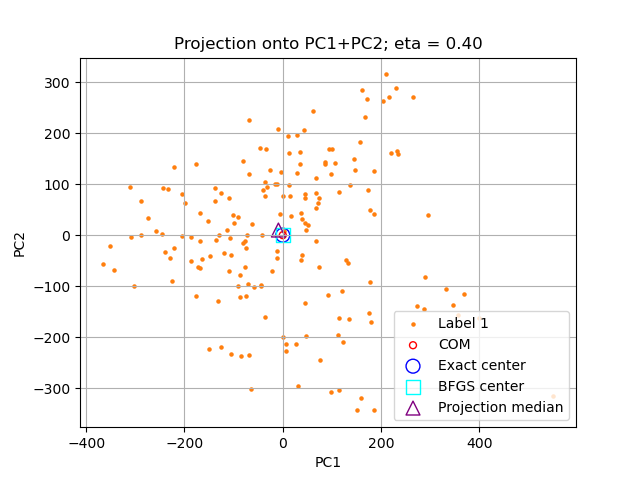} & \includegraphics[width=0.306\textwidth]{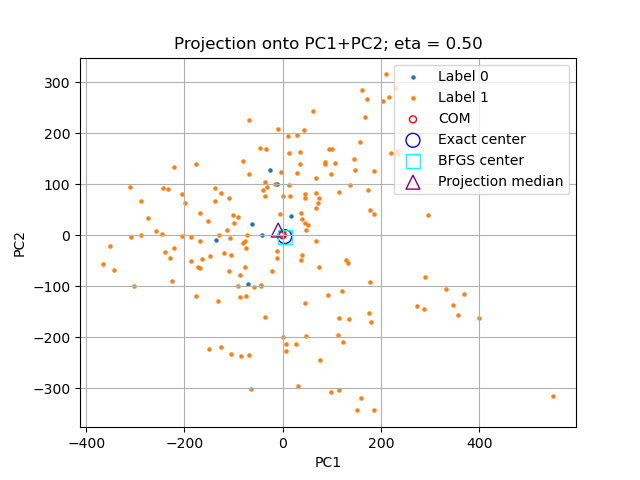} & \includegraphics[width=0.306\textwidth]{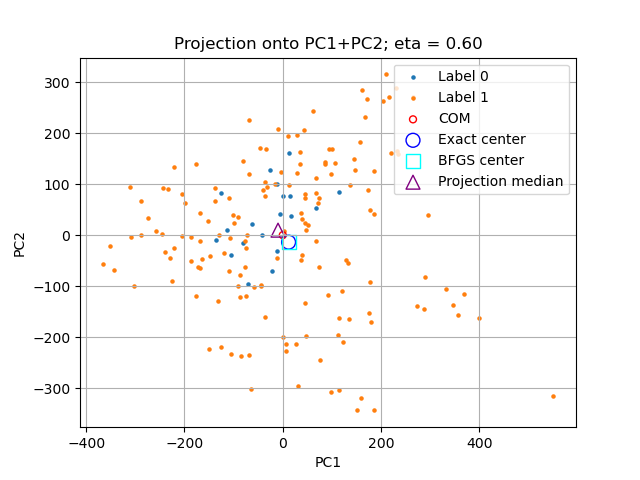}\\
{\scriptsize Projection plot cluster\_11-, $\eta=0.4$} &{\scriptsize Projection plot cluster\_11-, $\eta=0.5$} &Projection plot cluster\_11-, $\eta=0.6$\\
\includegraphics[width=0.306\textwidth]{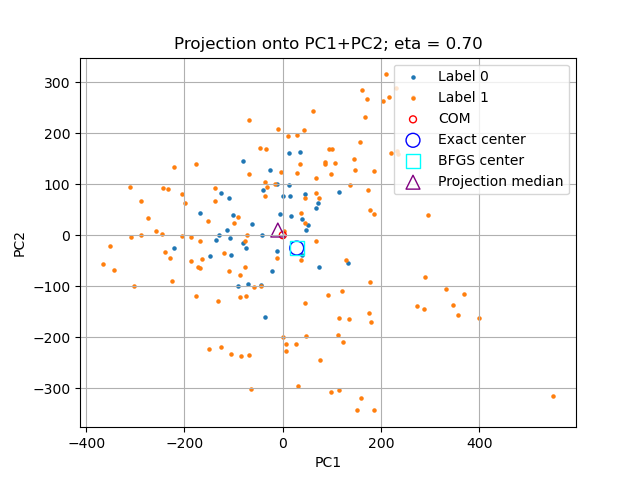} & \includegraphics[width=0.306\textwidth]{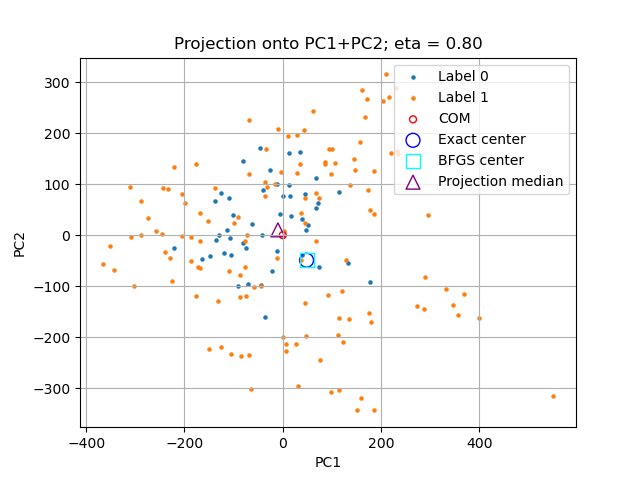} & \includegraphics[width=0.306\textwidth]{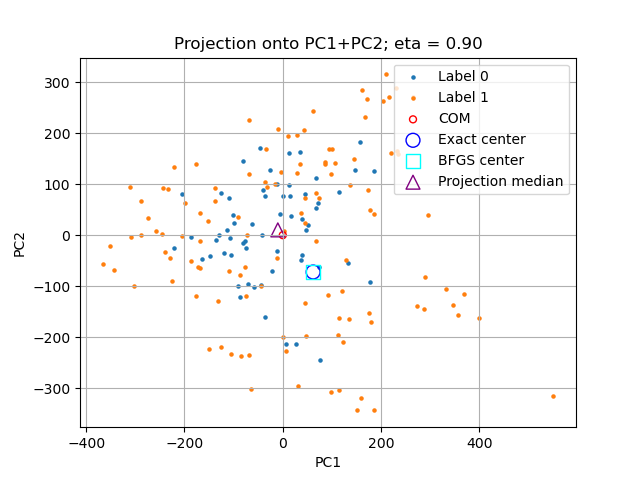}\\
{\scriptsize Projection plot cluster\_11-, $\eta=0.7$} &{\scriptsize Projection plot cluster\_11-, $\eta=0.8$} &{\scriptsize Projection plot cluster\_11-, $\eta=0.9$}
\end{tabular}
\end{center}
\caption{{\bf Dataset cluster\_11 -- shape (176,  400)}}
\label{fig:pp-cluster_11}
\end{figure}
\FloatBarrier

\subsubsection{Dataset Arcene}

\begin{table}[htb]
\begin{center}
\begin{tabular}{|l|rrr|}
\hline
 &  min  &  median  &  max\\
\hline
$\eta=0.10$ &  4.559e-05  &  4.559e-05  &  4.559e-05\\
$\eta=0.30$ &  5.732e-05  &  5.732e-05  &  5.732e-05\\
$\eta=0.50$ &  4.431e-05  &  4.431e-05  &  4.431e-05\\
$\eta=0.70$ &  5.946e-04  &  5.946e-04  &  5.946e-04\\
$\eta=0.90$ &  5.234e-02  &  5.234e-02  &  5.234e-02\\
\hline
\end{tabular}
\end{center}
\caption{{\bf DatasetHD: $\timeexact / \timelbfgs$.}}
\label{tab:DatasetHD-times-exact-LBFGS}
\end{table}

\begin{table}[htb]
\begin{center}
\begin{tabular}{|l|rrr|}
\hline
 &  min  &  median  &  max\\
\hline
$\eta=0.10$ &  1.000e+00  &  1.000e+00  &  1.000e+00\\
$\eta=0.30$ &  1.000e+00  &  1.000e+00  &  1.000e+00\\
$\eta=0.50$ &  1.000e+00  &  1.000e+00  &  1.000e+00\\
$\eta=0.70$ &  1.000e+00  &  1.000e+00  &  1.000e+00\\
$\eta=0.90$ &  9.447e-01  &  9.447e-01  &  9.447e-01\\
\hline
\end{tabular}
\end{center}
\caption{{\bf DatasetHD: $\Feta{\optexact} / \Feta{\optlbfgs}$.}}
\label{tab:DatasetHD-values-exact-LBFGS}
\end{table}

\FloatBarrier

\newpage
{\scriptsize
\tableofcontents
}

\end{document}